\documentclass[reprint,superscriptaddress,amsmath,amssymb,aps]{revtex4-2}

\usepackage{graphicx}
\usepackage{dcolumn}
\usepackage{bm}
\usepackage{lineno}
\usepackage{xcolor}
\usepackage{url}
\usepackage{bm}
\usepackage[breaklinks=true, colorlinks=true, linkcolor=blue]{hyperref}
\hypersetup{colorlinks=true, allcolors=blue, bookmarksopen=false, bookmarksnumbered}

\begin{document}

\title{Constraints on the onset of color transparency from quasi-elastic $^{12}$C$(e,e'p)$ up to $Q^2=\,14.2\,$(GeV$/c)^2$}
\newcommand*{\MSU }{Mississippi State University, Mississippi State, Mississippi 39762, USA}
\newcommand*{\MSUindex}{1}
\affiliation{\MSU}
\newcommand*{\UVA }{University of Virginia, Charlottesville, Virginia 22903, USA}
\newcommand*{\UVAindex}{2}
\affiliation{\UVA}
\newcommand*{\JLAB }{Thomas Jefferson National Accelerator Facility, Newport News, Virginia 23606, USA}
\newcommand*{\JLABindex}{3}
\affiliation{\JLAB}
\newcommand*{\REG }{University of Regina, Regina, Saskatchewan S4S 0A2, Canada}
\newcommand*{\REGindex}{4}
\affiliation{\REG}
\newcommand*{\NCAT }{North Carolina A \& T State University, Greensboro, North Carolina 27411, USA}
\newcommand*{\NCATindex}{5}
\affiliation{\NCAT}
\newcommand*{\KENT }{Kent State University, Kent, Ohio 44240, USA}
\newcommand*{\KENTindex}{6}
\affiliation{\KENT}
\newcommand*{\ZAG }{University of Zagreb, Zagreb, Croatia}
\newcommand*{\ZAGindex}{7}
\affiliation{\ZAG}
\newcommand*{\TEMP }{Temple University, Philadelphia, Pennsylvania 19122, USA}
\newcommand*{\TEMPindex}{8}
\affiliation{\TEMP}
\newcommand*{\YER }{A.I. Alikhanyan  National  Science  Laboratory \\ (Yerevan  Physics
Institute),  Yerevan  0036,  Armenia}
\newcommand*{\YERindex}{9}
\affiliation{\YER}
\newcommand*{\WM }{The College of William \& Mary, Williamsburg, Virginia 23185, USA}
\newcommand*{\WMindex}{10}
\affiliation{\WM}
\newcommand*{\CUA }{Catholic University of America, Washington, DC 20064, USA}
\newcommand*{\CUAindex}{11}
\affiliation{\CUA}
\newcommand*{\HU }{Hampton University, Hampton, Virginia 23669, USA}
\newcommand*{\HUindex}{12}
\affiliation{\HU}
\newcommand*{\FIU }{Florida International University, University Park, Florida 33199, USA}
\newcommand*{\FIUindex}{13}
\affiliation{\FIU}
\newcommand*{\CNU }{Christopher Newport University, Newport News, Virginia 23606, USA}
\newcommand*{\CNUindex}{14}
\affiliation{\CNU}
\newcommand*{\JAZ }{Jazan University, Jazan 45142, Saudi Arabia}
\newcommand*{\JAZindex}{15}
\affiliation{\JAZ}
\newcommand*{\UTENN }{University of Tennessee, Knoxville, Tennessee 37996, USA}
\newcommand*{\UTENNindex}{16}
\affiliation{\UTENN}
\newcommand*{\OHIO }{Ohio University, Athens, Ohio 45701, USA}
\newcommand*{\OHIOindex}{17}
\affiliation{\OHIO}
\newcommand*{\UCONN }{University of Connecticut, Storrs, Connecticut 06269, USA}
\newcommand*{\UCONNindex}{18}
\affiliation{\UCONN}
\newcommand*{\SBU }{Stony Brook University, Stony Brook, New York 11794, USA}
\newcommand*{\SBUindex}{19}
\affiliation{\SBU}
\newcommand*{\ODU }{Old Dominion University, Norfolk, Virginia 23529, USA}
\newcommand*{\ODUindex}{20}
\affiliation{\ODU}
\newcommand*{\ANL }{Argonne National Laboratory, Lemont, Illinois 60439, USA}
\newcommand*{\ANLindex}{21}
\affiliation{\ANL}
\newcommand*{\BOULDER }{University of Colorado Boulder, Boulder, Colorado 80309, USA}
\newcommand*{\BOULDERindex}{22}
\affiliation{\BOULDER}
\newcommand*{\ORSAY }{Institut de Physique Nucleaire, Orsay, France}
\newcommand*{\ORSAYindex}{23}
\affiliation{\ORSAY}
\newcommand*{\UNH }{University of New Hampshire, Durham, New Hampshire 03824, USA}
\newcommand*{\UNHindex}{24}
\affiliation{\UNH}
\newcommand*{\JMU }{James Madison University, Harrisonburg, Virginia 22807, USA}
\newcommand*{\JMUindex}{25}
\affiliation{\JMU}
\newcommand*{\RUTG }{Rutgers University, New Brunswick, New Jersey 08854, USA}
\newcommand*{\RUTGindex}{26}
\affiliation{\RUTG}
\newcommand*{\CMU }{Carnegie Mellon University, Pittsburgh, Pennsylvania 15213, USA}
\newcommand*{\CMUindex}{27}
\affiliation{\CMU}
\newcommand*{\VPI }{Virginia Polytechnic Institute and State University, Blacksburg, Virginia 24061, USA}
\newcommand*{\VPIindex}{28}
\affiliation{\VPI}

\author{D. Bhetuwal}\affiliation{\MSU}
\author{J. Matter}\affiliation{\UVA}
\author{H. Szumila-Vance}\affiliation{\JLAB}
\author{C.\,Ayerbe Gayoso}\affiliation{\MSU}\affiliation{\WM}   
\author{M.\,L.\,Kabir}\affiliation{\MSU}
\author{D. Dutta}\affiliation{\MSU}
\author{R.\,Ent}\affiliation{\JLAB}  

\author{D.\,Abrams}\affiliation{\UVA} 
\author{Z.\,Ahmed}\affiliation{\REG}  
\author{B.\,Aljawrneh}\affiliation{\NCAT}  
\author{S.\,Alsalmi}\affiliation{\KENT} 
\author{R.\,Ambrose}\affiliation{\REG} 
\author{D.\,Androic}\affiliation{\ZAG}  
\author{W.\,Armstrong}\affiliation{\ANL} 
\author{A.\,Asaturyan}\affiliation{\YER} 
\author{K.\,Assumin-Gyimah}\affiliation{\MSU}   
\author{A.\,Bandari}\affiliation{\WM}   
\author{S.\,Basnet} \affiliation{\REG}  
\author{V.\,Berdnikov}\affiliation{\CUA}    
\author{H.\,Bhatt}\affiliation{\MSU}   
\author{D.\,Biswas}\affiliation{\HU}\affiliation{\VPI}   
\author{W.\,U.\,Boeglin}\affiliation{\FIU} 
\author{P.\,Bosted}\affiliation{\WM} 
\author{E.\,Brash}\affiliation{\CNU}    
\author{M.\,H.\,S.\,Bukhari}\affiliation{\JAZ}  
\author{H.\,Chen}\affiliation{\UVA}             
\author{J.\,P.\,Chen}\affiliation{\JLAB}           
\author{M.\,Chen}\affiliation{\UVA}             
\author{E.\,M.\,Christy}\affiliation{\HU}          
\author{S.\,Covrig}\affiliation{\JLAB}           
\author{K.\,Craycraft}\affiliation{\UTENN}        
\author{S.\,Danagoulian}\affiliation{\NCAT}     
\author{D.\,Day}\affiliation{\UVA}             
\author{M.\,Diefenthaler}\affiliation{\JLAB}     
\author{M.\,Dlamini}\affiliation{\OHIO}          
\author{J.\,Dunne}\affiliation{\MSU}            
\author{B.\,Duran}\affiliation{\TEMP}            
\author{R.\,Evans}\affiliation{\REG}            
\author{H.\,Fenker}\affiliation{\JLAB}           
\author{N.\,Fomin}\affiliation{\UTENN}            
\author{E.\,Fuchey}\affiliation{\UCONN}           
\author{D.\,Gaskell}\affiliation{\JLAB}          
\author{T.\,N.\,Gautam}\affiliation{\HU}         
\author{F.\,A.\,Gonzalez}\affiliation{\SBU}       
\author{J.\,O.\,Hansen}\affiliation{\JLAB}           
\author{F.\,Hauenstein}\affiliation{\ODU}       
\author{A.\,V.\,Hernandez}\affiliation{\CUA}       
\author{T.\,Horn}\affiliation{\CUA}             
\author{G.\,M.\,Huber}\affiliation{\REG}       
\author{M.\,K.\,Jones}\affiliation{\JLAB}          
\author{S.\,Joosten}\affiliation{\ANL}          
\author{A.\,Karki}\affiliation{\MSU}            
\author{C.\,Keppel}\affiliation{\JLAB}           
\author{A.\,Khanal}\affiliation{\FIU}           
\author{P.\,M.\,King}\affiliation{\OHIO}             
\author{E.\,Kinney}\affiliation{\BOULDER}           
\author{H.\,S.\,Ko}\affiliation{\ORSAY}            
\author{M.\,Kohl}\affiliation{\HU}              
\author{N.\,Lashley-Colthirst}\affiliation{\HU}        
\author{S.\,Li}\affiliation{\UNH}               
\author{W.\,B.\,Li}\affiliation{\WM}               
\author{A.\,H.\,Liyanage}\affiliation{\HU}         
\author{D.\,Mack}\affiliation{\JLAB}              
\author{S.\,Malace}\affiliation{\JLAB}           
\author{P.\,Markowitz}\affiliation{\FIU}       
\author{D.\,Meekins}\affiliation{\JLAB}          
\author{R.\,Michaels}\affiliation{\JLAB}         
\author{A.\,Mkrtchyan}\affiliation{\YER}         
\author{H.\,Mkrtchyan}\affiliation{\YER}        
\author{S.J.\,Nazeer}\affiliation{\HU}         
\author{S.\,Nanda}\affiliation{\MSU}   
\author{G.\,Niculescu}\affiliation{\JMU}        
\author{I.\,Niculescu}\affiliation{\JMU}        
\author{D.\,Nguyen}\affiliation{\UVA} 
\author{Nuruzzaman}\affiliation{\RUTG}       
\author{B.\,Pandey}\affiliation{\HU}           
\author{S.\,Park}\affiliation{\SBU}             
\author{E.\,Pooser}\affiliation{\JLAB}           
\author{A.\,Puckett}\affiliation{\UCONN}         
\author{M.\,Rehfuss}\affiliation{\TEMP}          
\author{J.\,Reinhold}\affiliation{\FIU}         
\author{N.\,Santiesteban}\affiliation{\UNH}     
\author{B.\,Sawatzky}\affiliation{\JLAB}          
\author{G.\,R.\,Smith}\affiliation{\JLAB}             
\author{A.\,Sun}\affiliation{\CMU}               
\author{V.\,Tadevosyan}\affiliation{\YER}        
\author{R.\,Trotta}\affiliation{\CUA}           
\author{S.\,A.\,Wood}\affiliation{\JLAB}            
\author{C.\,Yero} \affiliation{\FIU}  
\author{J.\,Zhang}\affiliation{\SBU}        
\collaboration{Hall C Collaboration}
\noaffiliation

\date{\today}

\begin{abstract}
Quasi-elastic scattering on $^{12}$C$(e,e'p)$ was measured in Hall C at Jefferson Lab for space-like 4-momentum transfer squared $Q^2$ in the range of 8--14.2\,(GeV/$c$)$^2$ with proton momenta up to 8.3\,GeV/$c$. The experiment was carried out in the upgraded Hall C at Jefferson Lab. It used the existing high momentum spectrometer and the new super high momentum spectrometer to detect the scattered electrons and protons in coincidence. The nuclear transparency was extracted as the ratio of the measured yield to the yield calculated in the plane wave impulse approximation. Additionally, the transparency of the $1s_{1/2}$ and $1p_{3/2}$ shell protons in $^{12}$C was extracted, and the asymmetry of the missing momentum distribution was examined for hints of the quantum chromodynamics prediction of Color Transparency. All of these results were found to be consistent with traditional nuclear physics and inconsistent with the onset of Color Transparency.

\end{abstract}

\maketitle


\section{\label{sec:introduction}INTRODUCTION}
The $(e,e'p)$ reaction, also known as a proton-knockout reaction, is a fundamental tool for studying the propagation of nucleons in the nuclear medium. Specifically, the electromagnetic probe is able to sample the full nuclear volume (as compared to hadronic probes). The kinematics of the reaction are well-defined by the electron, and the momentum transferred can be independently varied from the energy transferred in the reaction. This enables a clean selection of parameter space for studying the propagation of the knocked-out proton through the nuclear medium and its final state interactions (FSI). The sensitivity to FSI makes quasi-elastic scattering an ideal probe of the phenomenon of Color Transparency (CT) predicted by Quantum Chromodynamics (QCD). 

Theoretical calculations in the quark-gluon framework of QCD predict that in exclusive processes at large, spacelike four-momentum transfer squared, $Q^2$, the FSI between the hadrons and the nuclear medium are reduced or suppressed. In the case of quasi-elastic electron scattering, only the FSI of the knocked-out proton are relevant. The concept of CT was first proposed by Mueller and Brodsky\,\cite{Mueller_1983, Brodsky_1982} in the context of perturbative QCD but was later shown to also arise in nonperturbative models. An analogue of CT can be seen in Quantum Electrodynamics: an $e^+e^-$ pair has a small interaction cross section near the production point acting as a dipole (neutral charge) instead of as isolated charged particles~{\cite{perkins, doi:10.1080/14786435808237003}}.

 The onset of CT requires the following conditions:
\begin{itemize}
    \item Squeezing: at sufficiently high $Q^2$, this is the preferential selection of a small configuration of quarks, sometimes referred to as a point-like configuration (PLC)
    \item Freezing: the PLC ejected at a high momentum maintains its small size over a distance comparable to or greater than the nuclear radius
    \item The in-medium interaction of the PLC as a color-neutral object is proportional to the square of its transverse radius 
    and thus, has reduced interaction with the nuclear medium as it transits the nucleus
\end{itemize}

Squeezing is experimentally controlled through the choice of the momentum transfer whereas freezing is described by the energy transfer of the reaction. It is the interplay between squeezing and freezing that is important to observing the onset of CT. 

The onset of CT has been observed in mesons \cite{elfassi12, ben_prl, xqian,  he4tranprc, Ackerstaff:1998wt, Arneodo:1994id}, whereas its onset in baryons remains uncertain with experimental results to date leading to ambiguous conclusions. For instance, the $pp$ scattering experiments at Brookhaven National Laboratory (BNL) \cite{Carroll:1988rp,Mardor:1998zf,Leksanov:2001ui} claimed to have initially found the onset of CT in protons, but the full results were inconsistent with a CT-only description. The BNL results have since been better explained with descriptions that include nuclear filtering~\cite{ralston} or exotic multi-quark final states~\cite{brodsky_ch}.


The nuclear transparency is the common observable for experiments searching for the onset of CT, and it is described as $T=\,\sigma_A/A\sigma_0$, or the ratio of the nuclear cross section per
nucleon, $\sigma_A/A$, to the cross section for a free nucleon, $\sigma_0$. Traditional Glauber multiple scattering theory \cite{vijay} predicts that $T$ is constant as $Q^2$ increases. It is specific to the qualities of QCD that one may predict the reduction of final state interactions, characterized as CT, subsequently resulting in an increase in the nuclear transparency with increasing $Q^2$.

All previous measurements of the momentum dependence of the nuclear transparency of protons (proton transparency) in quasi-elastic electron scattering have been consistent with the Glauber prediction, indicating no deviation with increasing momentum transfer. 
The most recent experiment, E12-06-107 - The Search for Color Transparency at 12 \,GeV \cite{E12-06-107}, took place at Jefferson Lab (JLab) and extended the range of $Q^2$ up to 14.2 (GeV/$c$)$^2$, the highest $Q^2$ studied to date for this reaction. The results indicate no signal consistent with the onset of CT ~\cite{PhysRevLett.126.082301} in this range. In this article we elaborate on the experimental details and report additional results on proton transparency separated by nuclear shells and the asymmetry of the missing momentum distribution.

\section{\label{sec:experiment}EXPERIMENTAL SET-UP}

This experiment was the first to be completed in Hall\,C after the beam energy upgrade of the continuous electron beam accelerator facility (CEBAF). The focus of this experiment was to study the semi-exclusive quasi-elastic $^{12}C(e,e'p)$ reaction, the knockout of a proton by an incident electron in a carbon target. 

The present experiment was designed to overlap with the existing $Q^2=\,8.1$\,(GeV/$c)^2$ data point from the highest $Q^2$ previous $A(e,e'p)$ measurements at JLab\,\cite{garrow} in order to help validate  the results. The present experiment measured nuclear transparency covering the range of outgoing proton momenta, $(p')$, of the BNL $\,A(p,2p)$ experiment where a rise in nuclear transparency had been previously reported~\cite{bnlfinal}. The use of an electron beam as opposed to a hadronic probe is ideal for such measurements as it avoids the ambiguity that arises from the reduction in flux of the probe when extracting the nuclear transparency. This measurement extended the $Q^2$ and $p'$ range to the highest achieved in quasi-elastic proton knockout to date.

Four kinematic settings were used in this experiment covering a range of $Q^2$=\,8--14.2\,(GeV/$c)^2$ and proton momenta from 5--8.3\,GeV/$c$. The kinematics for this experiment are shown in Table\,\ref{kinematic_table}. 

\begin{table}[hbt!]
 \caption{Kinematic settings of the experiment where $E_b$ is the electron beam energy, $p_{p}$ and $p_{\theta}$ correspond to the central momentum and angle of the proton spectrometer while  $e_{p}$ and $e_{\theta}$ correspond to the central momentum and angle of the electron spectrometer, and $\epsilon$ is the polarization of the virtual photon exchanged by the electron scattered at an angle $e_{\theta}$.}
    \label{kinematic_table}
\begin{center} 
\begin{tabular}{ccccccc} 
 \hline
 \rule{0pt}{2.5ex}
 $\textbf{E}_b$ & $\bm{Q^2}$ & $\textbf{p}_\theta$ & $\textbf{p}_p$ & $\textbf{e}_\theta$ & $\textbf{e}_p$ & $\epsilon$ \\ 
 \newline
 {(GeV)} & (GeV/$c)^2$ & (deg) & (GeV/$c$) & (deg) & (GeV/$c$) & $ $ \rule[-1.2ex]{0pt}{0pt}\\\hline
 $6.4$ & $8.0$ & $17.1$ & $5.030$ & $45.1$ & $2.125$ & $0.47$\\ 
 $10.6$ & $9.4$ & $21.6$ & $5.830$ & $23.2$ & $5.481$ & $0.76$\\ 
 $10.6$ & $11.4$ & $17.8$ & $6.882$ & $28.5$ & $4.451$ & $0.64$\\ 
 $10.6$ & $14.2$ & $12.8$ & $8.352$ & $39.3$ & $2.970$ & $0.44$\\ \hline
\end{tabular}
\end{center}
\end{table}

\subsection{Beam}

The experiment used the continuous wave (CW) electron beam with energies of 6.4 and 10.6\,GeV and beam currents of $10-65\,\mu$A. The electron beam is accelerated using superconducting radio frequency cavities. The duty factor of the beam is $\sim100\%$ and consists of pulses occurring at a frequency of $1497$\,MHz with an energy spread of $\pm 0.025\%$. The beam is sequentially delivered to all four experimental halls, allowing each experimental hall to operate simultaneously with different beam currents and energies\,\cite{harwood13}. Hall C received one out of three RF pulses from the accelerator, resulting in 499~MHz beam on the Hall C target. The beam energy was determined with an uncertainty of 0.1\% by measuring the bend angle of the beam on its way into Hall C while traversing a set of magnets with precisely known field integrals. 
    
 \subsection{Targets}
 
A 10\,cm long (726\,mg/cm$^2$) liquid hydrogen target was used for normalization to the elementary $ep$ scattering process. Two aluminum alloy foils placed 10\,cm apart were used to estimate the background from the end windows of the hydrogen target cell. The main production target was a carbon target of 4.9\% radiation lengths (rl), while a second carbon target of 1.5\% rl was used for systematic studies. The thicknesses of the targets were measured to better than 0.5\%. The beam incident on the liquid hydrogen target was rastered over a 2$\times$2\,mm$^2$ area to suppress density variations from localized boiling.
    
\subsection{Spectrometers}   

Hall C has two magnetic spectrometers, the High Momentum Spectrometer (HMS), which has been the main spectrometer in Hall C during the JLab 6\,GeV era, and the new Super High Momentum Spectrometer (SHMS). 
    
The HMS which served as the electron detection arm consists of three quadrupoles (Q) and a dipole (D) magnet arranged in a Q$_1$Q$_2$Q$_3$D configuration capable of bending the scattered particles vertically at an angle of 25$^{\circ}$ into the detector stack. The HMS has two available collimators of sizes approximately 8~msr and 4~msr; this experiment primarily used the larger collimator and compared the yields with the smaller collimator at a few select kinematic settings for systematic studies. Details about the HMS can be found in the Ref.~\cite{hms_info}. 
 
The SHMS which served as the proton detection arm has an extra dipole magnet known as the horizontal bender (HB) that bends the scattered particles horizontally by 3$^\circ$ from the beam line before reaching the first quadrupole. The configuration after the HB is the same as the HMS with three quadrupoles and the dipole magnet. The final dipole bends the particles by 18.4$^\circ$ vertically into the detector stack. The characteristics of both spectrometers are summarized in  Table\,\ref{tab:spectrometer_char}. 
 
The scattered electrons were detected in the HMS in coincidence with the knocked-out protons detected in the SHMS. The SHMS central angle was chosen to detect protons along the electron three-momentum transfer, $\vec{q}$. 
These kinematics minimize competing processes thereby simplifying the interpretation of any signal for the onset of CT. The measured final state proton momentum ranged from $5.030-8.352$\,GeV/$c$. The electron beam energy was 6.4\,GeV for the $Q^2$\,=\,$8.0$\,(GeV/$c)^2$ setting and 10.6\,GeV for the rest. 
    
   \begin{table}[hbt]
 \caption{Hall C Spectrometers characteristics}
    \label{tab:spectrometer_char}
\begin{center} 
\begin{tabular}{lrr} 
 \hline
   \noalign{\vskip 1px}                    & \textbf{HMS}\,\cite{donprl}  & \textbf{SHMS}\,\cite{shms_nim}  \\  \hline \noalign{\vskip 1px}  
    Momentum acceptance $\Delta p/p$ (\%)   & $\pm 10$      & -10 to +22 \\ 
    Solid angle acceptance $\Omega$ (msr)      & 8.1             & $>$4 \\ 
    Momentum resolution $(\%)$     & 0.1-0.15             & 0.03-0.08 \\ 
    Central momentum (p) \,(GeV/$c$)     & 0.4-7.4          & 2-11 \\ 
    Scattering angle $(\theta)\,(^\circ)$     & 10.5-90             & 5.5-40 \\
    Target position resolution (cm)     & 0.3             & 0.1-0.3 \\ 

  \hline

\end{tabular}
\end{center}
\end{table}
 
\subsection{Detectors} 
Each spectrometer in Hall C has a set of detectors stacked in the detector hut at the end of the spectrometer. Both spectrometers are equipped with a four-plane segmented hodoscope for triggering, time-of-flight measurements, and coarse tracking; multi-wire drift chambers for precision tracking; and a combination of a lead glass calorimeter and threshold Cherenkov counters for particle identification.

The HMS lead glass calorimeter and gas Cherenkov counter allow $e/\pi^-$ separation. The Cherenkov counter was filled with C$_4$F$_8$O at 0.45\,atm corresponding to an index of refraction of n=\,1.0006165 and a momentum threshold of 0.15\,GeV/$c$ for electrons and 3.97\,GeV/$c$ for pions. The HMS Cherenkov provides sufficient electron/pion discrimination for the highest and lowest kinematic points, but additional information from the calorimeter was required for the middle two kinematic points.

The SHMS is equipped with two gas Cherenkov detectors, one upstream and the other downstream of the drift chambers. Only the upstream Cherenkov detector was used in this analysis, and it was filled with CO$_2$ at 1 atm corresponding to an index of refraction of n=\,1.000449 with a momentum threshold of 4.66\,GeV/$c$ for pions and 31.1 \,GeV/$c$ for protons. 
The HMS and SHMS each contain pairs of drift chambers that give the hit position information of charged particles via the drift time for each hit that was used for track reconstruction. Two pairs of X-Y scintillator hodoscope planes in the HMS and SHMS formed the trigger for the data acquisition (DAQ). The fast timing response of the scintillators also measured the particle's time of flight (TOF) from the target. 
By using the particle track information from the drift chambers in combination with the timing information from the scintillators, the velocity of the particle ($\beta$) was determined and used to assist in particle identification. 

\section{Data Analysis}
\subsection{Calibrations}
\begin{figure*}[hbt!]
\centering
\includegraphics[width=\textwidth] {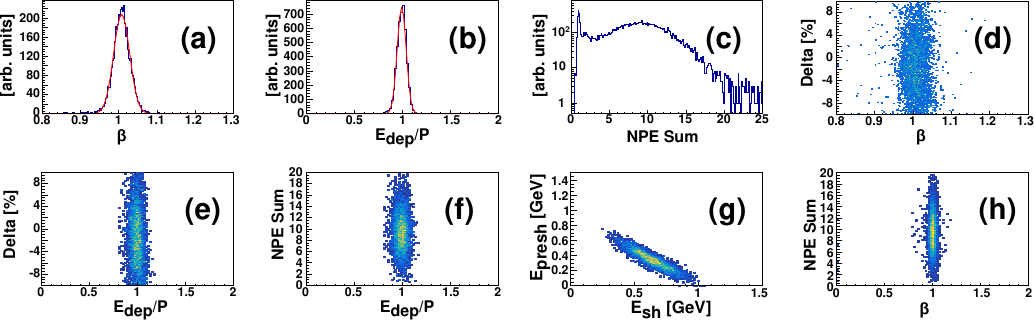}
\caption{Post-calibration response for hodoscopes, shower, preshower and calorimeter shown for the electron arm (HMS): (a) $\beta$, (b) E$_\textrm{tot}$/P (total energy deposited normalized by the central momentum), (c) total number of photoelectrons, (d) $\Delta$p /p vs $\beta$, (e) $\Delta$p /p vs E$_\textrm{tot}$/P, (f) total number of photoelectrons vs E$_\textrm{tot}$/P, (g) Shower Energy vs Preshower Energy, (h) total number of photoelectrons vs $\beta$}
 \label{cal_plots}
 \end{figure*}

The experiment used drift chambers, hodoscopes, Cherenkov detectors and calorimeters in both the HMS and SHMS. Each system was calibrated to match the signal arrival time for the individual scintillator elements and to match the gains of the calorimeter and Cherenkov signals. A few selected distributions from those calibrations are shown in Fig. \ref{cal_plots}.

The drift chamber calibration requires determining the start time offsets ($t_{0}$) on a per-wire basis. These $t_{0}$ offsets are the corrections by which the drift time spectrum of each wire must be shifted to ensure the start of the drift time distribution at 0~ns. For well-calibrated chambers, the distribution of drift distances (the distance an ionizing particle has to traverse across a cell) must be flat and the residual (the difference between the fitted track position determined from all planes and the hit location from an individual plane) distributions should have widths $\leq 250~\mu$m, corresponding to the tracking resolution for both the HMS and SHMS.  

The calibration of the calorimeters converts the digitized detector signal (i.e. output of the analog-to-digital converters (ADC)) into the total energy deposited by the particle. The calibration uses high statistics electron beam data and examines the normalized energy, defined as the energy deposited by the electron in the shower/preshower blocks in the calorimeter, divided by the momentum for all tracked charged particles. For a well calibrated calorimeter, this ratio peaks at unity with the minimum width possible and is independent of the relative momentum ($\delta$) and the position of the hit.

The hodoscopes provide the fast triggering and precise timing for the experiment. The timing calibration provides the timing correction value and is accomplished by determining the TOF offset and time walk corrections for each hodoscope paddle relative to a reference paddle in the stack. With the known offsets, the $\beta$ calculated from the TOF is peaked at unity independent of relative momentum, $\delta$, and the hit position. For more discussion on the detector calibration, see Ref.~\cite{bhetuwalthesis}.

\subsection{Beam charge accounting} 
The electron beam charge in Hall C is measured using several RF cavity Beam Current Monitors (BCMs) calibrated with an Unser parametric current transformer (PCT) having an extremely stable gain. The Unser is calibrated in situ by injecting a known current into a calibration wire. The Unser output signal is recorded against the known current. The slope of this linear relationship gives the gain. 
The Unser suffers from high noise and long term instability in the offset, but with sufficient integration and regular re-calibration of the offset, it can be used as an absolute beam current reference for the BCMs. 
The BCMs are stainless steel cylindrical waveguides that are tuned to the beam’s frequency (1497 MHz) and are designed for stable, low noise, non-destructive beam current measurements. As the electron beam passes through the cavity on its way to the target, it induces current in the cavity that is proportional to the intensity of the electron beam. The total accumulated beam charge was determined with $\approx 1$\% uncertainty. 

\subsection{Live time}    
 In order to calculate the experimental yield, it is necessary to consider those events arriving while the data acquisition (DAQ) is busy. This busy time reduces the overall live time, or availability, of the system to receive triggers. There are two main sources that reduce the system live time: the electronic-reduced live time from the period when the trigger hardware is busy, and the computer-reduced live time due to the finite time the DAQ computer needs to process and record events.

In this experiment the DAQ had a rate-dependent computer live time (CLT) which was  calculated from the ratio of recorded (accepted) physics triggers and the total physics triggers. To measure the live time due to all electronics modules in the DAQ system, an Electronics Dead Time Measurement (EDTM) trigger is inserted into the trigger logic. The EDTM rate was about $3 ~$Hz to minimize the probability of blocking actual physics triggers. The EDTM initiates a fake physics trigger to estimate the Total Live Time (TLT), which is calculated from the ratio of the number of EDTM triggers that are accepted by the DAQ to the total number of pulses counted by the EDTM scaler. 

The EDTM trigger was available during the experiment except for the lowest $Q^2$ of 8~(GeV/$c)^2$ setting. For this setting, we extrapolated from kinematics that had similar rates and a known live time. For more discussion on the live time calculations, see Refs.~\cite{carlosyerothesis,jmatterthesis}.

\subsection{\label{sect:optics} Spectrometer magnetic transport optimization}

The experiment was one of the first experiments to use the newly built SHMS to detect protons. The experiment used the SHMS over a wide range of central momenta and angles and measured the highest momentum protons in Hall C (8.3\,GeV/$c$) to date. Significant effort was made at the start of this experiment to characterize and optimize the SHMS magnetic transport of charge particles (optics). 
The fields for each of the magnets in the SHMS were modeled with the static field analysis code TOSCA~\cite{cite:TOSCA} and compared with field measurements. The Q$_2$ and Q$_3$ quadrupole magnets are nearly identical and have no saturation implemented in their models. The HB is characterized by a small degree of saturation above approximately 4\,GeV/$c$. The model for the HB magnet was compared against field mapping measurements along the central axis. The Q$_1$ magnet was also determined to have some saturation effects above approximately 7.5\,GeV/$c$, and these effects were measured only by measuring the central field values of the magnet versus the current to validate the more detailed TOSCA models. The magnets in the SHMS were set by their currents that were previously studied and validated with TOSCA models.

The HMS is generally well-understood through its extensive use in Hall C. The HMS analyzing dipole differs from that of the SHMS, as approximately half of its field is generated by the surrounding iron yoke of the magnet. As such, the HMS dipole is characterized by a larger settling time. The quadrupole magnets in the HMS were set using the same current to field ratios established and verified during previous use. The HMS spectrometer dipole is set by field regulation based on field values both measured and verified by TOSCA models. The well understood response of the HMS optics was further verified through hydrogen elastic measurements. 
\begin{figure}[hbt!]
\centering
\includegraphics[width=0.48\textwidth] {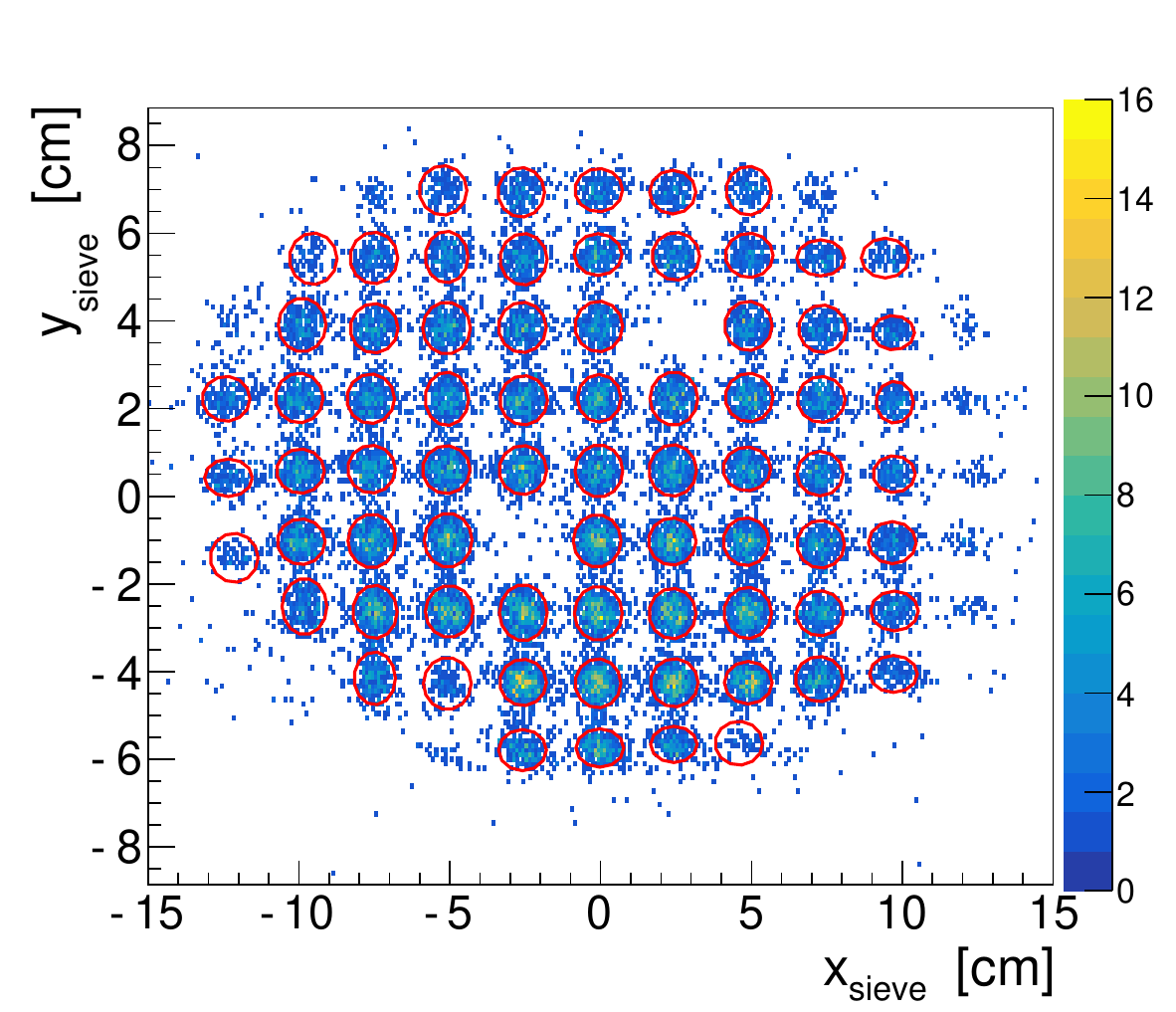}
 	     \caption{Reconstructed sieve aperture pattern for the central target foil in the SHMS. The central hole is half diameter compared to the other sieve holes (6~mm diameter), and two empty sieve positions are observed to be consistent with sieve holes that are blocked.}
 \label{fig:sieve}
 \end{figure}

Tracks reconstructed from the drift chamber hits provide the vertical (horizontal) position $x (y)$ and vertical (horizontal) angles $x'=\frac{dx}{dz} (y' = \frac{dy}{dz})$ of the particles at the focal plane located in between the two chambers. The positions and angles at the focal plane can be precisely mapped back to the position and angles at the interaction point in the target through a set of polynomial transformations. An initial set of coefficients for these transformations was generated using the COSY program\,\cite{Berz:2006zz}, which is a code for the simulation, analysis and design of particle optical systems, and is based on differential algebraic methods. The mapping was further optimized using dedicated data collected with a set of special purpose arrays of fixed apertures (sieve slits) and multi-foil extended carbon targets. The optics optimization data for both the HMS and SHMS were collected using the electron beam at an incident energy of 6.4\,GeV/$c$ with central spectrometer momenta of 2, 3, and 3.2\,GeV/$c$. Two targets were used to collect these data: a three-foil target with carbon foils at $\pm$10~cm and 0~cm, and a two-foil target with carbon foils at $\pm$5~cm along the beam direction ($z$). The sieve slits were placed downstream of the target in front of the first quadrupole magnet in each spectrometer arm. The events that passed through the sieve holes were used to optimize the reconstruction map using a singular value decomposition (SVD) algorithm~\cite{Press2007} to fine tune the coefficients generated from the COSY models and to accurately reproduce the positions and angles of the apertures. The optimized sieve aperture pattern for the SHMS is shown in Fig.\,\ref{fig:sieve}. 

The true sieve hole positions are shown by the grid intersections in Fig.\,\ref{fig:sieve}, and the events associated with those sieve holes are indicated by the red ellipse around those positions. The optimized mapping was valid up to central momenta of 3.2\,GeV/$c$. In the SHMS, there were some anticipated magnetic saturation effects in the horizontal bender and Q$_1$ magnets when the magnets were set for higher central momentum. These offsets were verified by observing the location of the waist of the focal plane distribution at these settings. 
The performance of the magnets at higher central momenta was fine-tuned by measuring the coincident elastic hydrogen reaction at each kinematic setting. There is no sieve data at the higher kinematic settings of this experiment to directly compare with the optimized optics that span up to a central momentum of 3.2~GeV$/c$. Nevertheless, magnet saturation and angle offset effects were well reproduced in simulation and yielded the correct reconstructed kinematics for the fully constrained H$(e,e'p)$ reaction. 

\subsection{Detector Efficiency}\label{sect:deteff}

Detector efficiency is defined as the ratio of the number of particles that passed threshold and fiducial cuts to the number of particles that traversed the detector and should have produced a signal in the detector under consideration. The calorimeter, Cherenkov and hodoscope efficiencies for the $^1$H and carbon targets were determined to be $\sim$\,99\% in both HMS and SHMS spectrometers. 

The tracking efficiency in the drift chambers is defined as the ratio of the number of events for which there was one track formed by the tracking algorithm to the number of events where one track was expected within a preselected region using the trigger scintillators. Variation in the tracking efficiency for the three independent preselected regions was used to determine the systematic uncertainty of the tracking efficiency. Tracking efficiency in the HMS spectrometer was found to be  $>$99\%, and in the SHMS spectrometer it ranged from 93\% $-$ 97\%. The tracking efficiency in the SHMS is rate dependent and is lower for the higher $Q^{2}$ corresponding to higher rates.
A series of dedicated single arm runs were taken on the carbon target to measure the charge normalized yield as a function of the beam current (also known as a luminosity scan). 
Within measurement uncertainties, it is expected that the corrected, charge-normalized carbon yield should be independent of beam current. The uncertainties due to the live time correction, and the detector and trigger inefficiencies were determined from a set of luminosity scans performed with each spectrometer at the beginning and at the end of the experiment. The charge normalized yield from these scans for each spectrometer was found to be independent of the beam current within statistical uncertainties, and the average variation in the normalized yield vs beam current was recorded as the systematic uncertainty, which we determined to be 0.5\%. 

\subsection{Target density reduction}
The density of the 10\,cm liquid $^1$H target can vary with the incident electron beam current (at a microscopic level as the $e^-$ beam interacts with the target, the number of target atoms in a local unit volume changes as the beam deposits power on it), and the experimental yield was corrected for this effect. The nominal liquid $^1$H pressure was 165~kPa with a temperature of 19~K. A series of dedicated single arm runs at different beam currents were taken to study the density reduction effect in the $^1$H target before and after collecting the production data. The charge normalized yield was determined as a function of the beam current. A linear fit of the reduction in yield as a function of the increasing beam currents was used to obtain a target density reduction correction to all of the experimental yields. The correction was determined to be 2.6\% at the highest beam current used, which was 65\,$\mu$A. 

\subsection{Simulation of the Experiment}
\subsubsection{Acceptance}

The acceptance of the spectrometers was studied using the SIMC simulation tool\,\cite{simc}. SIMC includes models generated by COSY for the spectrometer optics that transport the charged particles through the magnetic fields of all magnets in each spectrometer arm. The effects of multiple scattering and ionization energy loss for particles passing through all materials and apertures is included in the forward transport simulation. A second set of maps generated by COSY is used to relate the particle tracks at the focal plane of the spectrometer to the angles, momentum, and position at the interaction vertex in the target. Simulated events are weighted by the calculated Plane Wave Impulse Approximation (PWIA) cross-section, radiative correction, and Coulomb correction. The PWIA cross-section was calculated using the De Forest\,\cite{deForest} $\sigma_{cc1}$ prescription for the off-shell electron-proton cross-section and an independent particle shell model (IPSM) spectral function for the target nucleus~\cite{ne18_1}.

\begin{figure*}[hbt!]
\centering
\includegraphics[width=\textwidth]{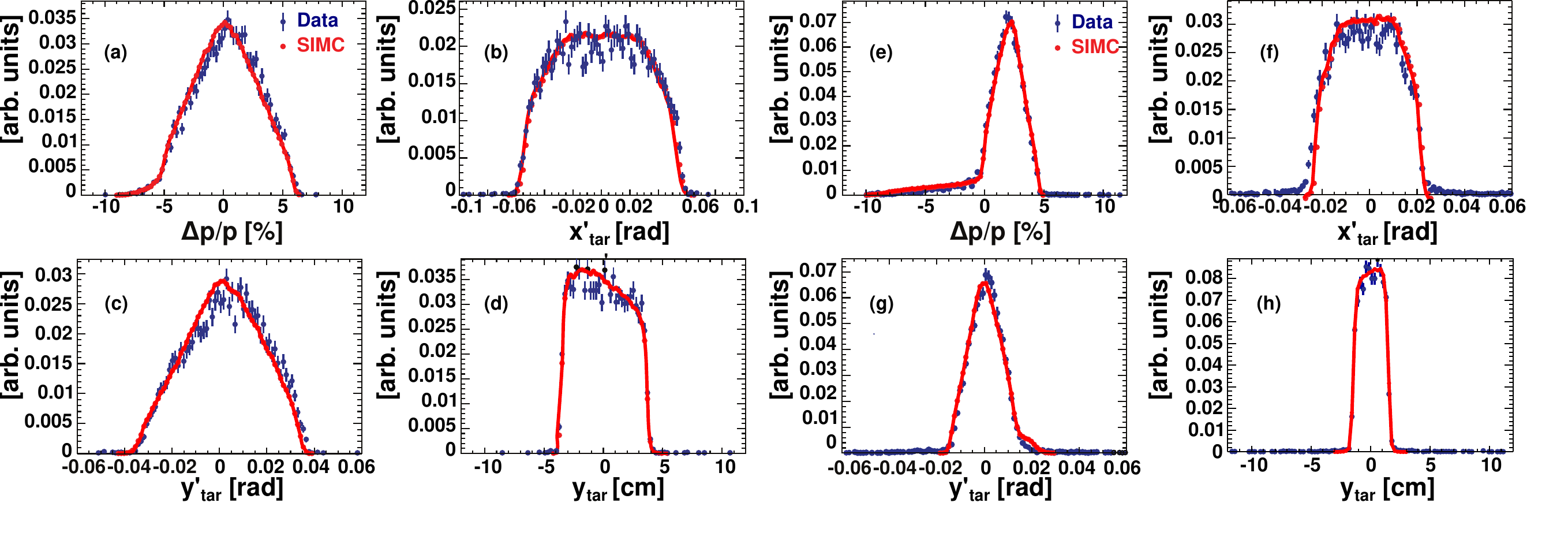}
   \caption[]{The reconstructed angles at the target and momentum for data (blue) and simulated spectra (red) for the measured H$(e,e'p)$ reaction with arbitrary normalization. Panels (a) - (d) show  the momentum bite $\Delta p/p$ (a), vertical angle ($x'_\textrm{tar}$) (b), horizontal angle ($y'_\textrm{tar}$) (c) and reconstructed horizontal position ($y_\textrm{tar}$) for the electrons in the HMS. Panels (e) - (g) show the momentum bite $\Delta p/p$ (e), vertical (f), horizontal angle (g), and reconstructed horizontal position (h) for the proton in the SHMS.}
  \label{fig:acc_heep}
\end{figure*}

The reconstructed angles and momentum at the target from coincident hydrogen elastic scattering obtained from simulation are compared to data in Fig.\,\ref{fig:acc_heep}. The exclusive nature of elastic scattering was used to better validate the spectrometer optics and to ultimately quantify how well the true acceptance is modeled. As a typical example, the comparisons between data and SIMC for the $Q^2$=\,8\,(GeV/$c)^2$ kinematics are shown in Fig.\,\ref{fig:acc_heep}.    
\noindent The yield from the SIMC simulation was obtained by accounting for the experimental luminosity, the phase space volume, and the number of events generated.

\subsubsection{Spectral functions}

The PWIA $(e,e'p)$ differential cross-section can be written as the product of $ep$ cross-section $(\sigma_{ep})$ and a probability function $S(E_s,\vec{p}_m)$, also known as the~spectral~function:
\begin{equation}
    \frac{d^6\sigma}{dE_{e'}d\Omega_{e'}dE_{p'}d\Omega{p'}} =\, p'E_{p'}\sigma_{ep}S(E_s,\vec{p}_m),
    \label{eq:pwia_cross_section}
\end{equation}
where $E_{e'}$ is the energy of the scattered electron, $E_{p'}$ is the energy of the knocked out proton, $p'$ is the measured outgoing proton momentum, and $\Omega_{e'}$, $\Omega_{p'}$ are the solid angles of the outgoing electron and proton respectively. The spectral function represents the probability of measuring a proton with missing momentum ${\textbf{p}}_m$ and separation energy $E_s$ (experimentally measured as missing energy, $E_m$). The two quantities ${\vec{p}}_m$ and $E_m$ are defined as:
\begin{equation}
 \vec{p}_m = \vec{p'}-\vec{q}, {\mbox{~~and~~}}  E_m = \nu - T_{p} - T_{A-1},\\
\end{equation}
where $\vec{q}$ and $\nu$ are the momentum and energy transferred between the incident and scattered electron respectively, $T_{p}$ is the kinetic energy of the struck proton and $T_{A-1}$ is the kinetic energy of the (undetected) recoiling $A-1$ system. In our experiment, we work in parallel kinematics such that $\vec{p'}$ is parallel to $\vec{q}$. 

In the IPSM, the nucleons are treated as free particles, and the spectral function has a different probability for each shell. However, it neglects that the nucleons are bound and hence off-shell. This means $E^2 \ne \vec{p}\,^2+M^2$, in general, where $E$, $p$, and $M$ are the energy, momentum, and mass of the bound nucleon, respectively. The electron scattering cross-section depends on the proton's initial energy, which yields two alternatives, either $E=\,M-E_s$ or $E^2=\,\vec{p}\,^2+M^2$. The choice of assumptions results in different off-shell cross-section prescriptions. 

The two often-used off-shell prescription models are De Forest $\sigma_{cc1}$ and $\sigma_{cc2}$ \cite{PhysRevC.53.2304, DEFOREST1983232}. The subscript $cc$ refers to the current conservation, and obeys $\vec{q}\,\vec{J}=\,\nu\rho$, with $\vec{q}$ the virtual photon three momentum, $\vec{J}$ the nuclear current density, $\nu$ is the virtual photon energy, defined before, and $\rho$ the nuclear charge density. This experiment uses the De Forest $\sigma_{cc1}$ prescription for the off-shell cross-section. The full computed cross-section model for all kinematics was observed to be insensitive to the choice of off-shell prescription (between $\sigma_{cc1}$ and $\sigma_{cc2}$) at $<0.1$\%. The IPSM spectral functions used in previous experiments\,\cite{ne18_1,ne18_2,donprl,garrow} were employed in this experiment.

\subsubsection{Radiative corrections}

Electrons radiate in the presence of nuclei or other electrons. In electron scattering experiments  this radiation results in an unwanted background in the spectrum of the scattered electrons. These so-called radiative tails must be accurately accounted for in order to extract any reliable information from the experimental spectra. Mo and Tsai\,\cite{RevModPhys.41.205} developed a comprehensive formulation for a set of approximations that could be used to correct a wide range of electron scattering processes. The radiative corrections in the SIMC simulation were based on this formulation adapted for the coincidence $(e,e'p)$ reaction~\cite{entradcor}. 

\begin{figure}[hbt!]
\centering
\includegraphics[width=0.45\textwidth] {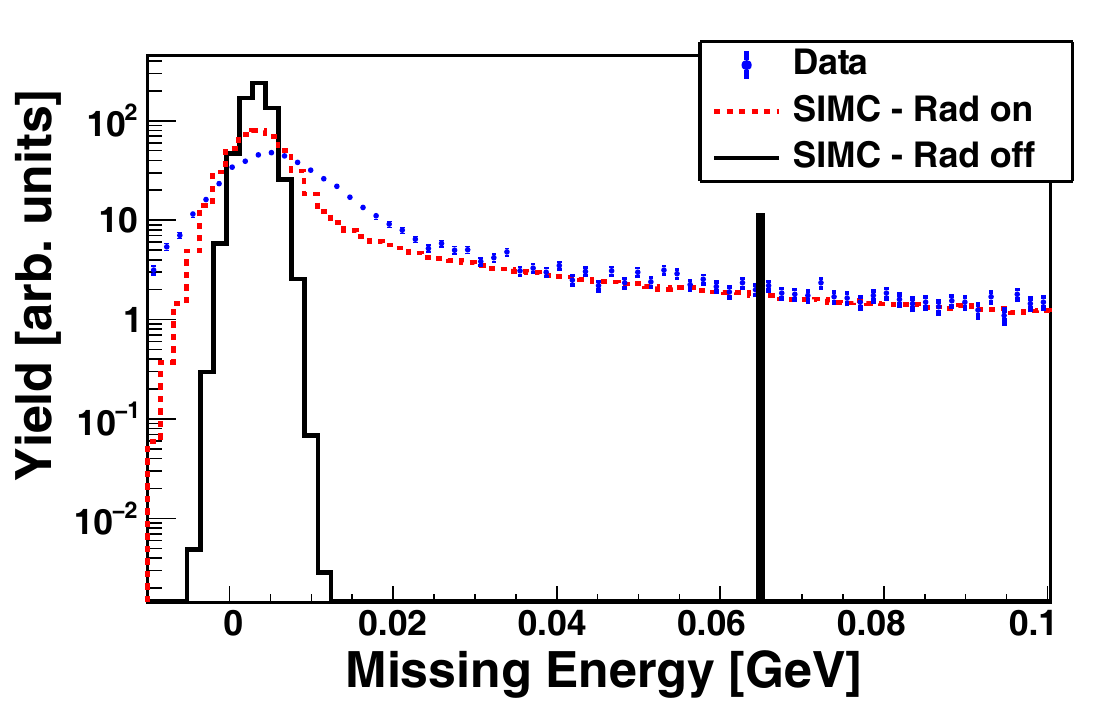}
 	     \caption{Hydrogen missing energy spectra for $Q^2$=\,8\,(GeV/$c)^2$ comparing data (blue dots) and Monte Carlo with (red dashed line) and without (black line) radiative correction. The vertical black line at 65 MeV indicates the $E_{\textrm{miss}}$ cut for hydrogen.}
 \label{q8_lh2_rad}
 \end{figure}

\begin{figure}[hbt!]
\centering
\includegraphics[width=0.45\textwidth] {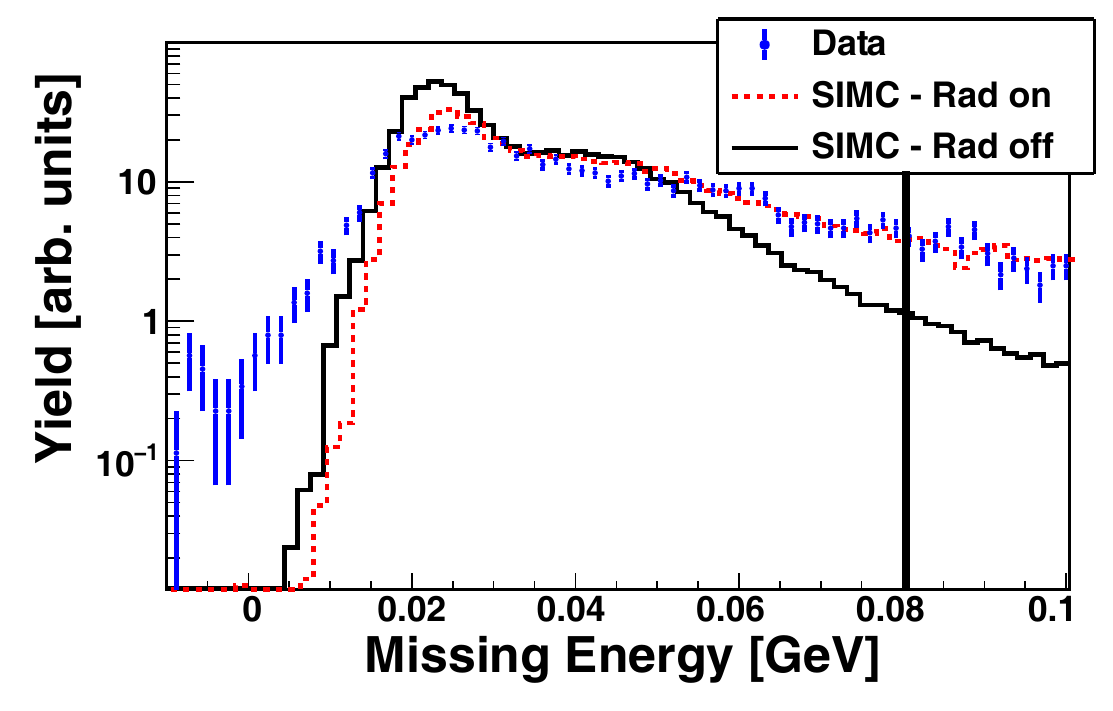}
 	     \caption{Carbon missing energy spectra for $Q^2$=\,8\,(GeV/$c)^2$ comparing data (blue dots) and Monte Carlo with (red dashed line) and without (black line) radiative correction. The vertical black line at 80 MeV indicates the $E_{\textrm{miss}}$ cut for carbon.}
 \label{q8_c12_rad}
 \end{figure}

Figures \ref{q8_lh2_rad} and \ref{q8_c12_rad} are the hydrogen and carbon missing energy distributions for $Q^2$= 8 (GeV/c)$^2$, respectively. In both figures, the data and Monte Carlo distributions are compared. Also shown are the locations of the missing energy cuts applied to both data and Monte Carlo: 65 MeV for hydrogen and 80 MeV for carbon. The sharply peaked solid black distributions and the broadened red dashed distributions show the Monte Carlo without and with radiation, respectively. The high missing energy tails seen in the data distributions are well reproduced by the simulation when radiation is included.

\subsection{Proton Absorption}
\label{sec:pa}
Because protons are strongly interacting particles, they may undergo a nuclear reaction as they pass through the materials of the SHMS before forming a trigger. The proton absorption, $A$, is defined as the fraction of protons that fail to form a trigger due to their interaction in the matter between the target and the detectors. An  estimation of the absorption is obtained by considering the proton's mean free path in the materials along its trajectory through the SHMS from:
\begin{itemize}
    \item the nuclear collision length: $\lambda_{T}=\,\sum_i A_i / (N_A \rho_i \sigma_{\textrm{tot}_i})$ where $N_A$ is Avogadro's number, $A_i$ the atomic weight, $\rho_i$ the mass density and $\sigma_{\textrm{tot},i}$ the total nuclear cross-section of the $i$th component of the material composition. 
\item the nuclear interaction length: $\lambda_{I}$, which is similarly defined as the nuclear collision length but subtracts the elastic and quasi-elastic cross-sections from $\sigma_{\textrm{tot},i}$.
\end{itemize}

Because the elastic cross-section is peaked in the forward direction, thus removing only a few protons from the spectrometer's acceptance, we use the average $\bar\lambda$ of $\lambda_T$ and $\lambda_I$ as our estimate of the mean free path. The estimated absorption is $A=\,1-e^{-\sum_i l_i/\bar\lambda_i}~\sim~\,8\%$ where $l_i$ is the thickness of each material in the proton's path. The collision and interaction lengths were taken from the PDG\,\cite{pdg_material_properties}, which are independent of the proton momenta in the momentum range of this experiment.

The proton absorption estimated using the mean-free-path was validated by comparing the charge-normalized coincident yield ($Y_{\textrm{coin}}$) and electron-only yield ($Y_{\textrm{sing}}$) recorded in the HMS for hydrogen elastic  $^1$H$(e,e'p)$ runs. The $Y_{\textrm{sing}}$ was obtained for a small central region of HMS acceptance along with tight limits  on the invariant mass $W$ ensuring a clean sample of electrons that participated in elastic scattering. $Y_{\textrm{coin}}$ was obtained with the same tight limits on the HMS acceptance and provided the yield for detected protons. The proton absorption given by $A=\,1-Y_{\textrm{coin}}/Y_{\textrm{sing}}$ is the fraction of events where an elastic electron event in the HMS did not produce a corresponding proton in the SHMS. Using the $Q^2$=\,11.5\,(GeV/$c)^2$ data, we obtain a proton absorption of $A=\,9.0 \pm 0.7 \%$. The uncertainty quoted here is the quadrature sum of the statistical uncertainty and a systematic uncertainty estimated by varying the cuts used to calculate yields. The two methods used to estimate the proton absorption are consistent with each other within uncertainty. 
The difference between the two methods (1\%) added in quadrature with the uncertainty of the data driven method (0.7\%) was used to obtain the overall systematic uncertainty due to the proton absorption quoted in Table~\ref{errors}.  

\subsection{Systematic Uncertainty}
The systematic uncertainties are categorized into two sources: $Q^2$-dependent uncertainty (which includes uncertainty due to spectrometer acceptance, event selection, tracking efficiency, radiative corrections, live time and detector efficiency) and normalization uncertainty (which includes uncertainty due to the $ep$ cross-section, target thickness, beam charge, and proton absorption). Table\,\ref{errors} lists the major sources of systematic uncertainties, and the sum in quadrature of these two sets of uncertainties is $4.0\%$. 
Since $\vec{p}_m$ relies on the momentum and angle reconstruction for both of the spectrometers, it is the most sensitive variable to validate the quality of the spectrometer acceptance model. The acceptance uncertainty was determined by quantifying the differences in the shape of the $|\vec{p}_m|$ distribution between data and SIMC, and was found to be $\sim$~2.6\%. The systematic uncertainty arising from the cut dependence of the experimental yield was determined by varying the cuts one at a time and recording the variation in yields for the different kinematic settings and the targets. The quadrature sum of the variation over all the different cuts was used as the event selection uncertainty, which we determined to be 1.4\%. The tracking efficiency was continuously monitored with an uncertainty of about 0.1\% for the HMS and $<$ 0.5\% for the SHMS. The uncertainty in the tracking efficiency was obtained from the average variation of the SHMS tracking efficiency when using the three independent methods for determining the efficiency (see Section~\ref{sect:deteff}). The uncertainty due to radiative corrections was estimated by comparing the tail of the missing energy spectra from the 1.5\% radiation length carbon data, and varying the $E_m$ cut. 
The measured $ep$ elastic cross-section with the hydrogen target with the background from the aluminum target cell subtracted, agrees with the world data. A comparison to a Monte Carlo simulation yields an overall normalization uncertainty of 1.8\%. 

\begin{table}[htb!]
\caption{\label{errors} Systematic Uncertainties}
\centering
\begin{tabular}{lc}\hline 
\rule{0pt}{2.5ex}
Source & $Q^2$ dependent uncertainty (\%) \\ \hline
\rule{0pt}{2.5ex}Spectrometer acceptance & 2.6\\
Event selection & 1.4 \\
Tracking efficiency & 0.5\\
Radiative corrections & 1.0 \\ 
Live time \& Detector efficiency  & 0.5 \\ 
\hline
\rule{0pt}{2.5ex}
Source & Normalization uncertainty (\%) \\ \hline
\rule{0pt}{2.5ex}Elastic $ep$ cross-section & 1.8 \\
Target thickness & 0.5\\
Beam charge & 1.0\\
Proton absorption & 1.2\\\hline
\rule{0pt}{2.5ex}Total & 4.0\% \\ \hline
    \end{tabular}
\end{table} 

The thicknesses of the carbon targets were measured to better than 0.5\% which is taken as the systematic uncertainty due to target thickness. The variation in the charge-normalized experimental yield was $<$1\% when using all events with beam current above 5\,$\mu$A or a more restrictive cut of $\pm$\,3\,$\mu$A around the average current (for each interval with stable current). This validates the $\sim$\,1\% uncertainty assigned to the beam charge measurement. 
 
\section{\label{sec:results}RESULTS}

\subsection{Hydrogen elastics}

The coincident elastic scattering reaction from the hydrogen target, H$(e,e'p)$, was used to fully constrain the spectrometer optics models used to reconstruct the momentum and angle, to fully understand detector efficiencies, and to determine the overall charge-normalized yield. This exclusive scattering reaction was measured at all four kinematic settings of the experiment (see Table~\ref{kinematic_table}). In elastic $ep$ scattering, the reconstructed invariant mass $W$ is most sensitive to the electron kinematics measured by the HMS. The offset between the reconstructed $W$ and the expected $W$ was primarily accounted for, at all kinematic settings, by offsets or imperfections in setting the central momentum and angle of the spectrometer. These offsets vary with each setting of the HMS central momentum. The HMS central momentum was offset by as much as 0.4\% at the highest central momentum (corresponding to the largest offset with respect to $W$ of approximately 60~MeV) due to magnet saturation effects. 

Due to the generally very large energy transfers to the proton, the missing energy and missing momentum are strongly correlated to the proton kinematics measured by the SHMS. Offsets in the central momentum and optics of the SHMS were improved by studying the focal plane dependencies of the residual difference of the reconstructed missing energy and the missing energy as calculated without the proton information. From simulations with slightly mistuned magnets, it was observed that first order corrections to the polynomial transformation coefficients (see Sect.~\ref{sect:optics}) were sufficient to remove the dependency of such residuals and was consistent with the offset of the magnet tune mis-sets. 

The yields from hydrogen scattering were used to determine how well the overall normalization of the data was understood. The missing energy and missing momentum cuts on the elastic hydrogen data were varied from 40 to 80\,MeV. The average deviation in the ratio of the charge-normalized yield to the simulation was determined to be no greater than 1\%. 

\begin{figure}[hbt!]
 \centering
  \includegraphics[width=0.49\textwidth]{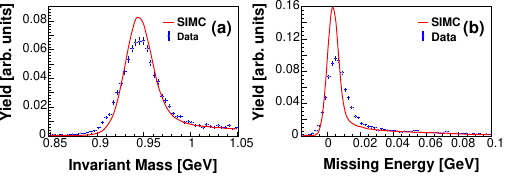}
 \caption[]{ The comparison between simulation and data through the H$(e,e'p)$ reaction (with arbitrary normalization) is shown for the $Q^2$=\,8\,(GeV$/c$)$^2$ setting. The reconstructed $W$ (a) is primarily driven by the electron arm (HMS) reconstruction, while the missing energy (b) includes contributions from the proton arm (SHMS).}
  \label{fig:H_q8_plots}
\end{figure}

The reconstructed $W$ and missing energy for hydrogen scattering is shown in Fig.\,\ref{fig:H_q8_plots} for the $Q^2$=\,8\,(GeV$/c)^2$ kinematic setting. 
Some additional resolution effects can be observed in the widths of the distributions relative to the simulated spectra. 
The reconstructed $W$ and missing energy peak locations show generally good agreement with simulation, and the high missing energy tail agrees well with simulation where contributions due to radiative effects are dominant. 

We constructed the ratios between the measured hydrogen elastic yields and the yields expected from simulation for $E_{\textrm{m}}<$65\,MeV and $|\vec{p}_{\textrm{m}}|<$ 65\,MeV$/c$. These cuts were varied in increments of 5\,MeV (5\,MeV$/c$) over the range of 40--80\,MeV (40--80\,MeV$/c$) for $E_{\textrm{m}}$ ($p_{\textrm{m}}$).
The average deviation of the ratios at each setting was found to be no greater than 1\%. A comparison between the ratios at the $Q^2$=\,9.5\,(GeV$/c$)$^2$ setting when the small and large collimators were used indicated a maximum deviation of 1.5\% between the yields. These uncertainties, combined, account for a 1.8\% uncertainty on the measured hydrogen elastic cross-section. For the four kinematic settings, the ratio of the hydrogen elastic data yield to simulation was unity.

\subsection{Transparencies}

In constructing the transparency, the ratio of the carbon yield is compared to the yield predicted from PWIA simulation. 
The measured carbon yield is first corrected for the detector-related inefficiencies. 

The carbon yields in both data and simulation were cut at $E_{\textrm{m}}<0.08$\,GeV and $p_{\textrm{m}}<0.3$\,GeV$/c$. For these cuts in carbon, the effect of nucleon-nucleon (NN) short-range correlations was previously determined to shift the single-particle strength to higher $p_{\textrm{m}}$, (i.e. some protons are shifted to higher $p_{\textrm{m}}$ due to short-range interactions with other nucleons) requiring a correction factor to be applied to the data (same factor for all kinematic settings) of 1.11 $\pm$ 0.03\,\cite{ne18_1}. This cut and the corresponding correction factor were used in the previous experiments \cite{ne18_1, garrow, donprl, ne18_2} and are independent of $Q^2$. The total model-dependent uncertainty of 3.9\% includes uncertainty in the spectral function (2.8\%) and the nucleon-nucleon correlation effects~\cite{ne18_1}.

The simulated yield is calculated for the same phase-space volume as the experiment. The carbon transparency was observed to be independent of $Q^2$ from 8--14.2\,(GeV$/c$)$^2$ ruling out observations that would be consistent with the onset of CT~\cite{PhysRevLett.126.082301} in this range.  

\subsection{Nuclear shell dependent transparency}
\begin{figure}[hbt!]
\centering
\includegraphics[width=0.49\textwidth] {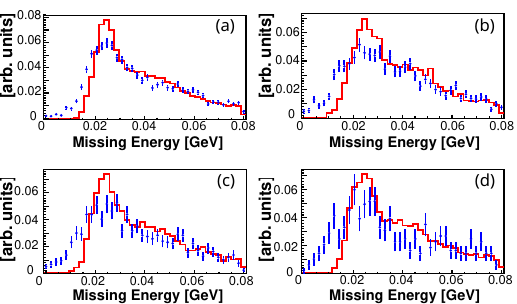}
\caption[Carbon missing energy spectrum for the experiment data.]{ Carbon missing energy spectra for the experimental data (blue points) for each of the 4 kinematic settings in $Q^2$: (a) 8, (b) 9.4, (c) 11.4, and (d) 14.4\,(GeV$/c$)$^2$ compared to simulation for the corresponding kinematics (red line). These spectra include both the $1s_{1/2}$ and $1p_{3/2}$ shell contributions.}
\label{em_plot}
\addtocontents{lof}{\vspace{\baselineskip}}
\end{figure}

In the $^{12}C(e,e'p)$ reaction, the protons knocked out from different nuclear shells (for example the $1s_{1/2}$ and $1p_{3/2}$ shells) are expected to have measurable differences in their attenuation by the nuclear medium. These differences arise from the differences in the intrinsic momentum distributions of protons occupying different nuclear shells, the differences in quenching of the nuclear shell occupation probabilities, and the presence of a hole around the struck proton due to short-range NN repulsion~\cite{frankfurt1990color}. These effects should lead to differences in the measured nuclear transparency. In addition, Frankfurt \textit{et\,al.} \cite{frankfurt1990color} suggests that the reduction of FSI (i.e. the CT effect) is more prominent for the $1s_{1/2}$ protons than in $1p_{3/2}$ protons due to differences in the soft re-scattering contributions to the hole excitation. They conclude that it may be advantageous to measure the ratio of the nuclear transparency of protons knocked out of the $1s_{1/2}$ and  $1p_{3/2}$ shells, as many experimental errors and theoretical
uncertainties are likely to cancel out, making the ratio a more sensitive probe of CT.
\begin{figure}[hbt!]
\centering
\includegraphics[width=0.49\textwidth] {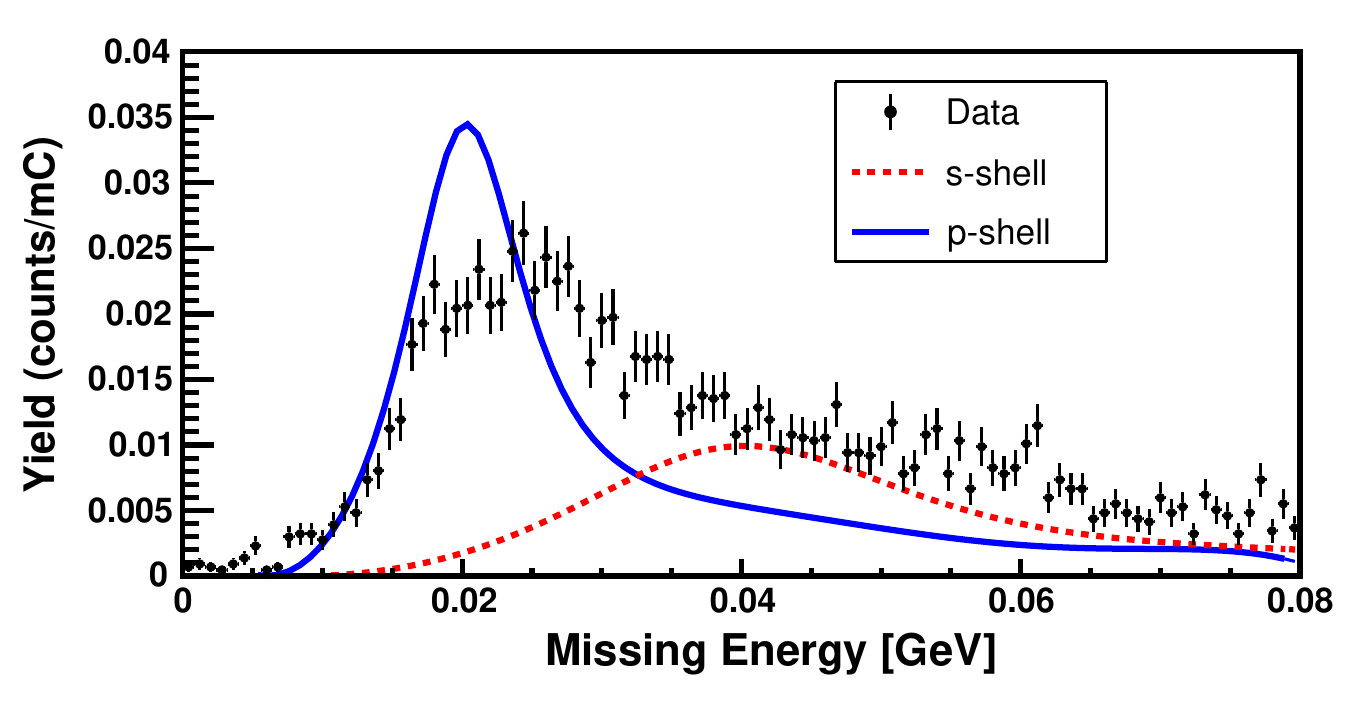}
\caption{The solid blue (dashed red) distribution is the simulated $1p_{3/2}$ ($1s_{1/2}$) shell contribution fitted to a Lorentzian distribution and a polynomial background. The black points with error bars (statistical only) are the data distribution from the corresponding $Q^2$=\, 8\,(GeV/$c)^2$ kinematics.}
\label{q8_111}
\addtocontents{lof}{\vspace{\baselineskip}}
\end{figure}
\begin{figure}[hbt!]
\centering
\includegraphics[width=0.49\textwidth] {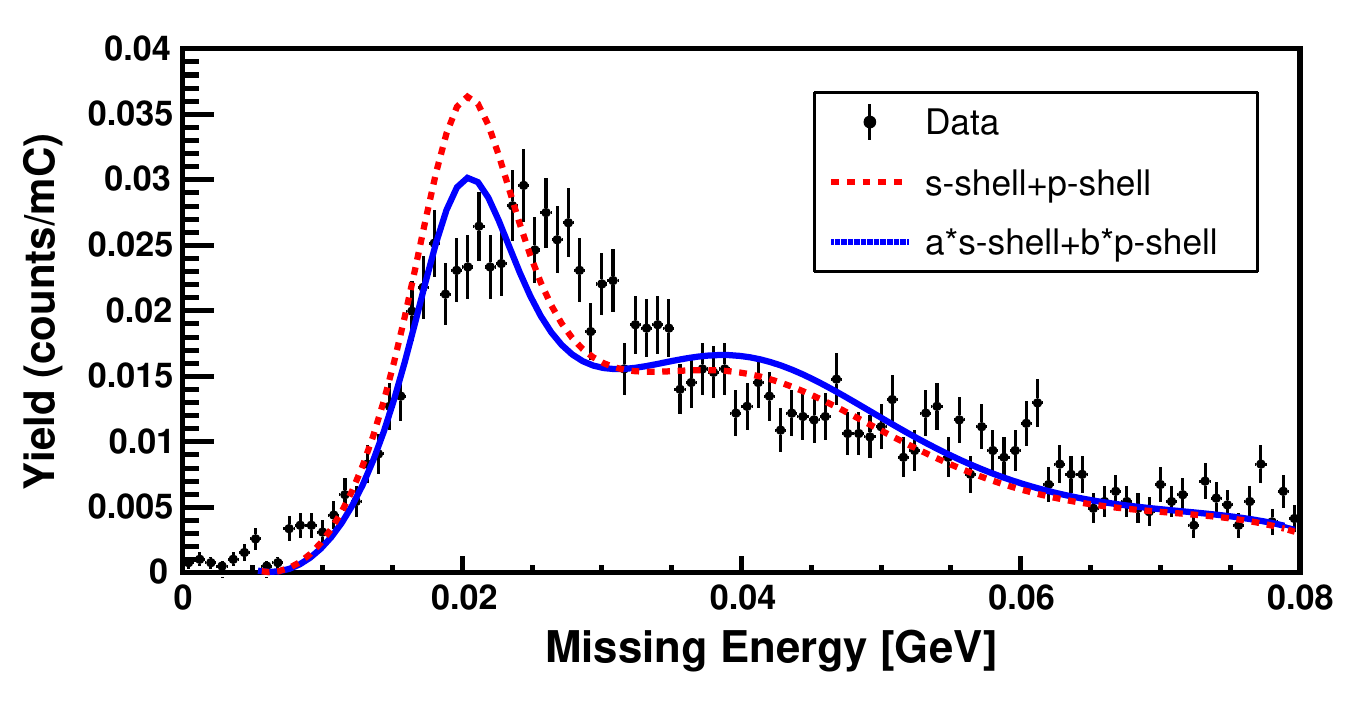}
\caption{The dashed red distribution is the sum with unit weights of the $1s_{1/2}$ shell and $1p_{3/2}$ contributions, the solid blue distribution is the 
$a\,(1s_{1/2})+ b\,(1p_{3/2})$ distribution for best fit to the data as described in the text. The black points with error bars are the data (statistical errors only). All the distributions correspond to $Q^2$=\, 8\,(GeV/$c)^2$ kinematics.}
\label{q8_222}
\addtocontents{lof}{\vspace{\baselineskip}}
\end{figure}

 In order to distinguish the $1s_{1/2}$ shell and $1p_{3/2}$ shell protons (higher and lower missing energy respectively), the data are shown as a function of the missing energy in Fig.~\ref{em_plot} for each kinematic setting. Also shown are the simulated missing energy distributions.
The reconstructed missing energy resolution is insufficient at these high $Q^2$ kinematics (due to the resolution of the high momentum protons) to cleanly separate the $1s_{1/2}$ and $1p_{3/2}$ shell contributions.
Therefore, instead of using a single excitation energy to separate the different shell contributions, we have adopted a simulation-driven method. The simulated contributions from the $1s_{1/2}$ and $1p_{3/2}$ shells, fitted to a Lorentzian distribution and a polynomial background, are shown separately in Fig.~\ref{q8_111} along with the data. These fits had a reduced $\chi^2$ ranging from 0.8 -- 2.1 for the different Q$^2$ values. The simulation uses the constraint that the carbon nucleus has 2 protons in the $1s_{1/2}$ shell, and 4 protons in the $1p_{3/2}$ shell. The simulation also uses a constant nuclear transparency normalization factor of 0.56 for the carbon target. All extracted transparencies are relative to this normalization factor of 0.56. The Lorentzian fits to the simulated  $1s_{1/2}$ and $1p_{3/2}$ shell spectra were then parameterized as $a\,(1s_{1/2})+ b\,(1p_{3/2})$,  and the best fit values for the parameters $a$ and $b$ were obtained by fitting to the measured yield. The sum with unit weights of the $1s_{1/2}$ and $1p_{3/2}$ contributions (red dashed line) compared to data is shown in Fig.~\ref{q8_222}. The combined distribution for the parameters obtained from the best fit to the data is shown as the blue solid distribution. These fits had a reduced $\chi^2$ ranging from 1.0 -- 2.9 for the different Q$^2$ values.\\
\begin{figure}[hbt!]
\centering
\includegraphics
[width=0.49\textwidth] 
{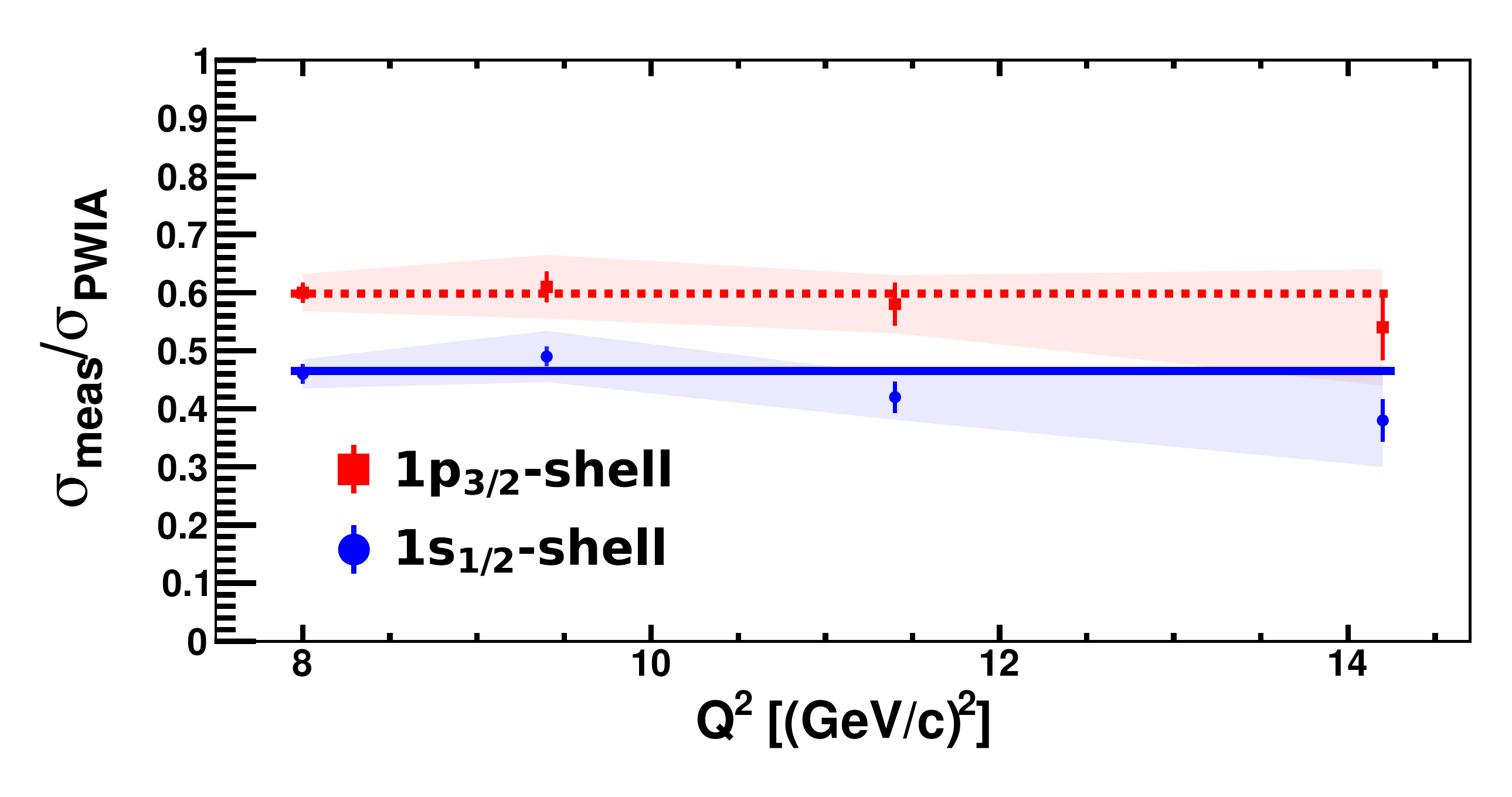}
\caption{ $1s_{1/2}$ (blue circles) and $1p_{3/2}$  shells (red squares) transparency as a function of $Q^2$. The straight lines are fit to a constant value for the respective shells. The error bars on each point show the statistical uncertainty while the bands represent the total systematic uncertainty of the $1p_{3/2}$ shell (red), and $1s_{1/2}$ shell (blue) transparencies. Note that there is an additional 3.9\% model-dependent uncertainty that is not shown in the figure.}
\label{shell_t_plot}
\addtocontents{lof}{\vspace{\baselineskip}}
\end{figure}
\begin{figure}[hbt!]
\centering
\includegraphics[width=0.49\textwidth] 
{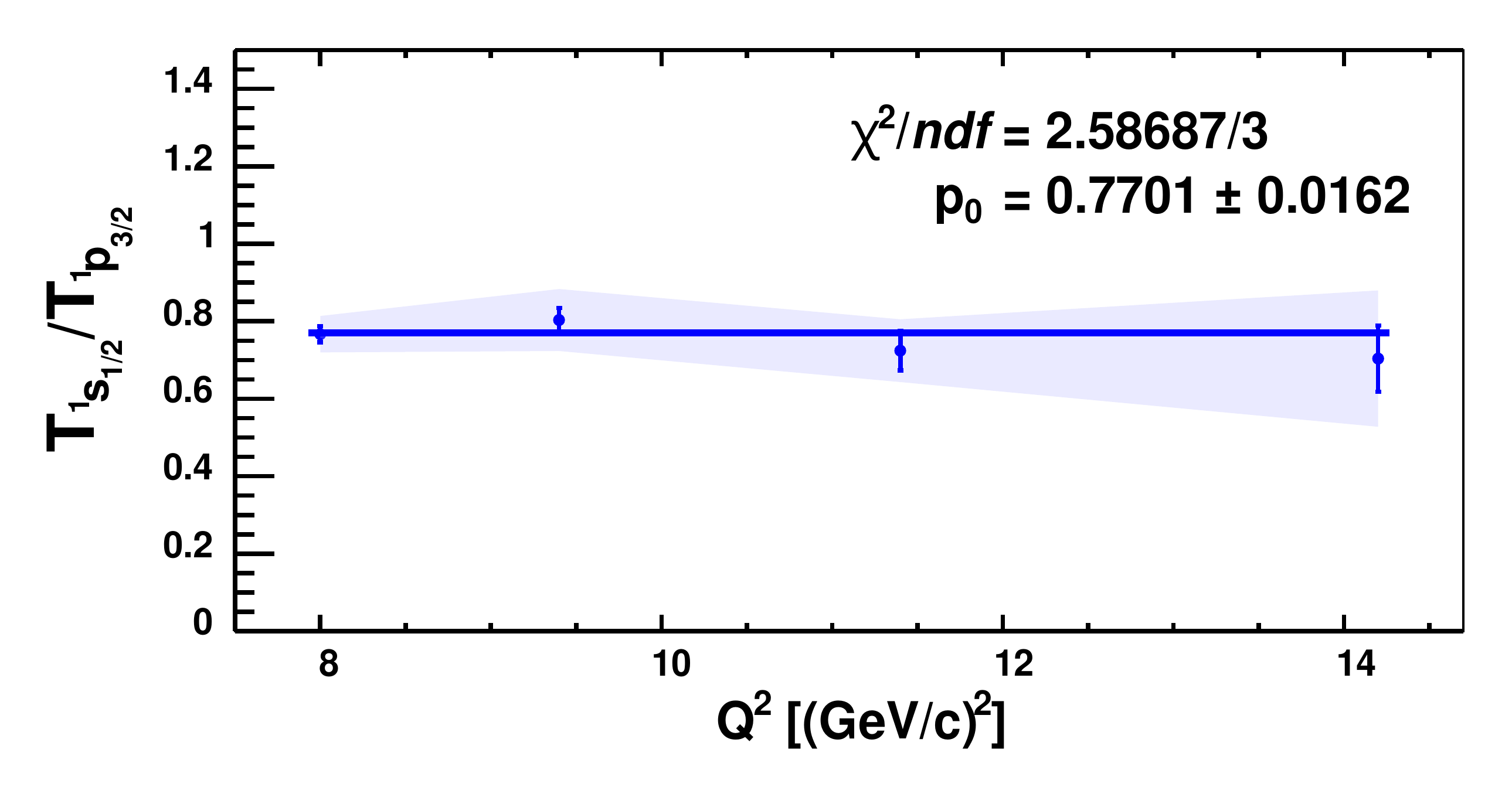}
\caption{The ratio of transparencies for the $1s_{1/2}$ shell to the $1p_{3/2}$ shell protons as a function of $Q^2$. The error bars show the statistical uncertainty, while the band represents the total systematic uncertainty. The solid line shows the fit to a constant value.}
\label{stop_ratio}
\addtocontents{lof}{\vspace{\baselineskip}}
\end{figure}
\begin{table*}[htb!]
\caption{\label{errors_carbon_1p_result} The $1s_{1/2}$ and $1p_{3/2}$  shell transparencies for the $^{12}C$ nucleus along with statistical, systematic and total uncertainties.}
\addtocontents{lot}{\vspace{\baselineskip}}
\centering

\begin{tabular}{cccccccccc}\hline 
& \multicolumn{4}{c}{$1s_{1/2}$} & & \multicolumn{4}{c}{$1p_{3/2}$}\\
             \cline{2-5}   \cline{7-10}
$Q^2$ & Transparency & Statistical & Systematic & Total & & Transparency & Statistical & Systematic & Total \\ 
(GeV/$c)^2$& $(T)$  & error & error & \% & & $(T)$  & error & error & \% \\
\hline
8.0  & 0.46 & 0.01 & 0.03 & 5.89 & & 0.60 & 0.01 & 0.03 & 5.58\\
9.4  & 0.49 & 0.01 & 0.04 & 9.12 & & 0.61 & 0.02 & 0.05 & 9.53\\
11.4 & 0.42 & 0.02 & 0.04 & 10.37 & & 0.58 & 0.03 & 0.05 & 10.14\\
14.2 & 0.38 & 0.03 & 0.08 & 22.23 & & 0.54 & 0.05 & 0.10 & 21.03\\ \hline
\end{tabular}
\end{table*} 

The nuclear transparency of the $1s_{1/2}$ and $1p_{3/2}$ shell protons is obtained from the product of normalization factor and the parameters $a$ or $b$. The $1s_{1/2}$ and $1p_{3/2}$ shell transparencies for each $Q^2$ are listed in Table~\ref{errors_carbon_1p_result}. The total systematic uncertainty for $1s_{1/2}$ and $1p_{3/2}$ shell transparencies includes the uncertainty of the fit parameters and the normalization uncertainty and are summarized in Table \ref{errors_carbon_1p_result}.
\begin{table}[htb!]
\caption{\label{fit_param_final} Results of the fit to a constant transparency as a function of $Q^2$ for the combined, $1p_{3/2}$ and $1s_{1/2}$ shells transparencies.}
\centering
\begin{tabular}{c c c c c c c}
\hline
Fit result & \hspace{0.5cm} & combined &\hspace{0.5cm}  & $1p_{3/2}$ shell  &\hspace{0.5cm}  & $1s_{1/2}$ shell \\ 
\hline
$\chi^2/df$ &  &2.08 &  &0.70 & &6.53 \\
$T_{\textrm{fit}}$ & &0.56$\pm$0.01 & &0.60$\pm$0.01 & &0.46$\pm$0.01 \\
\hline
\end{tabular}
\end{table} 

The shell-dependent transparency as a function of $Q^2$ is shown in the Fig.~\ref{shell_t_plot}. The blue and the red bands are the systematic uncertainties, which are the quadrature sum of the $4\%$ systematic uncertainty and the uncertainty of determining the $1s_{1/2}$ shell and $1p_{3/2}$ shell transparencies separately. 
The shell-dependent 
transparencies were also fit to a constant value, with the constant values and the quality of the fits listed in Table~\ref{fit_param_final}.
The shell-dependent nuclear transparency shows little variation with $Q^2$ and does not show the onset of CT-like behavior.

The ratio of the nuclear transparency from $1s_{1/2}$ to $1p_{3/2}$ shell is shown in Fig.~\ref{stop_ratio}. The differences between the $1s_{1/2}$ and $1p_{3/2}$ shell transparencies arise from the differences in the momentum distributions, excitation energy and differences in the re-distribution of strength due to nucleons in short-range correlations, radiative effects and the presence of a hole around the struck proton due to short-range NN repulsion. 
The possible cancellation of experimental and theoretical uncertainties makes the ratio of the $1s_{1/2}$ to $1p_{3/2}$ shell transparencies a more sensitive observable of CT compared to the transparency averaged over the two shells. The onset of CT would be observed as an increase in the ratio with increasing $Q^2$. However, as can be seen in Fig.~\ref{stop_ratio}, the transparency ratio is independent of $Q^2$ reinforcing the observed lack of CT-like effects at the kinematics probed in this experiment. \\

\subsection{Asymmetry of the missing momentum distribution}
\begin{figure}[hbt!]
\centering
\includegraphics[width=0.49\textwidth] {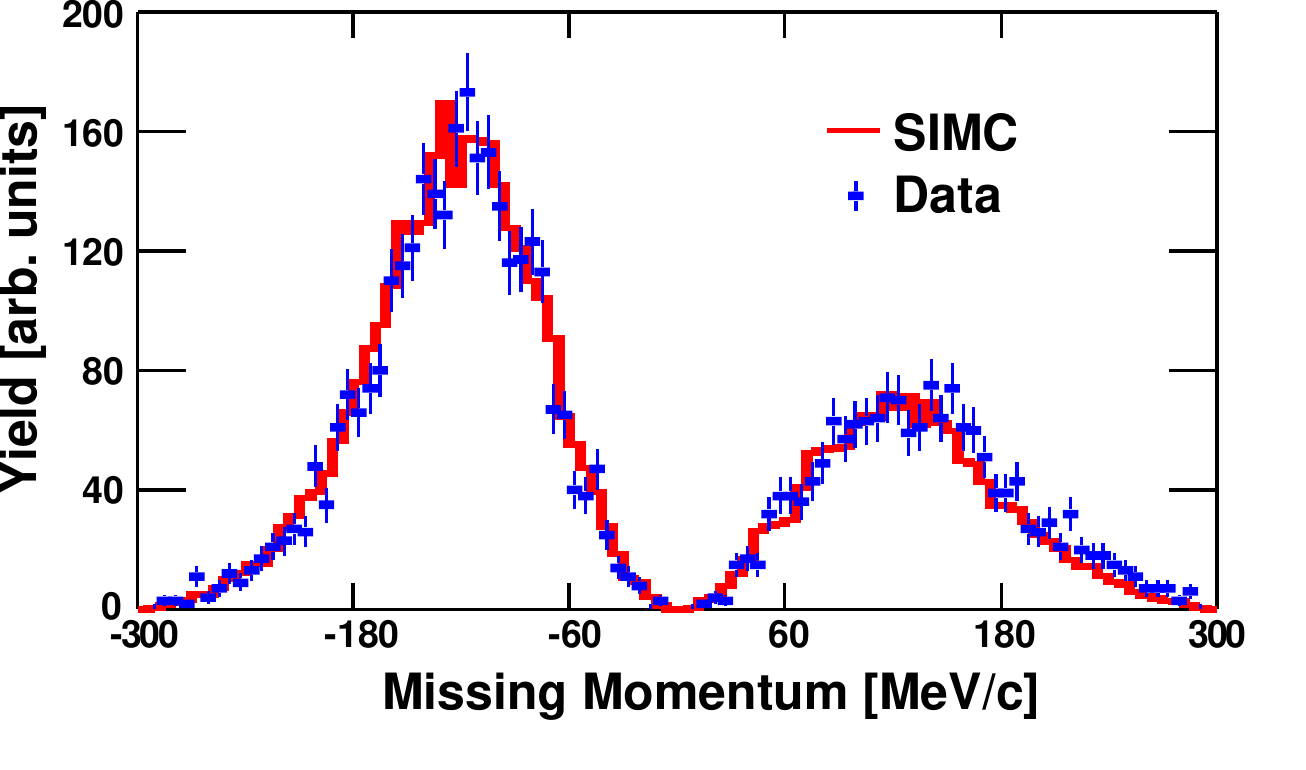}
\caption{The missing momentum distribution is shown for the kinematic setting at $Q^2=8$~(GeV/$c)^2$ where the simulation is normalized to the data.}
\label{fig:pm}
\end{figure}

 In parallel kinematics under the PWIA, the distribution of events with the missing momentum $\vec{p}_m$ parallel (negative) and anti-parallel (positive) to the direction of momentum transfer $\vec{q}$ is symmetric. The differences in the experimental acceptance for negative and positive $p_m$ give rise to most of the asymmetry that is observed in the missing momentum spectrum as shown in Fig.~\ref{fig:pm}. 
A small fraction of the asymmetry is due to the small but finite angular coverage of protons on the left and right side of $\vec{q}$. This asymmetry is modified by FSI mechanisms beyond the impulse approximation including Meson Exchange Currents (MEC) and Isobar Configurations (IC)~\cite{PhysRevC.70.034606, PhysRevLett.83.5451}. Further, it was suggested that the Fermi motion of bound nucleons may be a source of CT in quasielastic scattering, particularly when the initial momentum of the bound nucleon is in the direction opposite to $\vec{q}$~\cite{jennings93}. This implies that CT is highly dependent on the sign of $p_m$~\cite{Bianconi:1993mol}. This is because all the excited baryon states are produced preferentially at positive $p_m$, and therefore, it is more probable to realize a point-like-state for positive $p_m$. 
 
 Therefore, it is interesting to measure the $Q^2$ dependence of the missing momentum asymmetry. This asymmetry, $A_{p_m}$, can be quantified as
\begin{equation}
    A_{p_m} = \frac{N_+ -N_-}{N_++N_-}
\end{equation}
with $N_+$ being the number of events integrated over a fixed range of positive $p_m$ and $N_-$ being the number of events integrated over the same range of negative $p_m$. The $p_m$ and $E_m$ dependence of $A_{p_m}$ was studied by dividing the $p_m$ range of $\pm \leq$~300 MeV/c into 5 equal bins with $E_m \leq $~80 MeV for each bin and the $E_m\leq$~80 MeV range into 4 equal bins with $\pm \leq$~300 MeV/c for each bin, respectively. This ensures that we exclude the regions where the impulse approximation is invalid and could influence the asymmetry from sources other than quasi-elastic scattering. The systematic uncertainty due to the binning in $p_m$ and $E_m$ was determined to be 11\% by varying the exact bin boundary. The PWIA simulation of the experiment can describe the $\vec{p}_m$ asymmetry very well as seen in Fig.~\ref{fig:pm}. 
\begin{figure}[hbt!]
\centering
\includegraphics[width=0.45\textwidth] 
{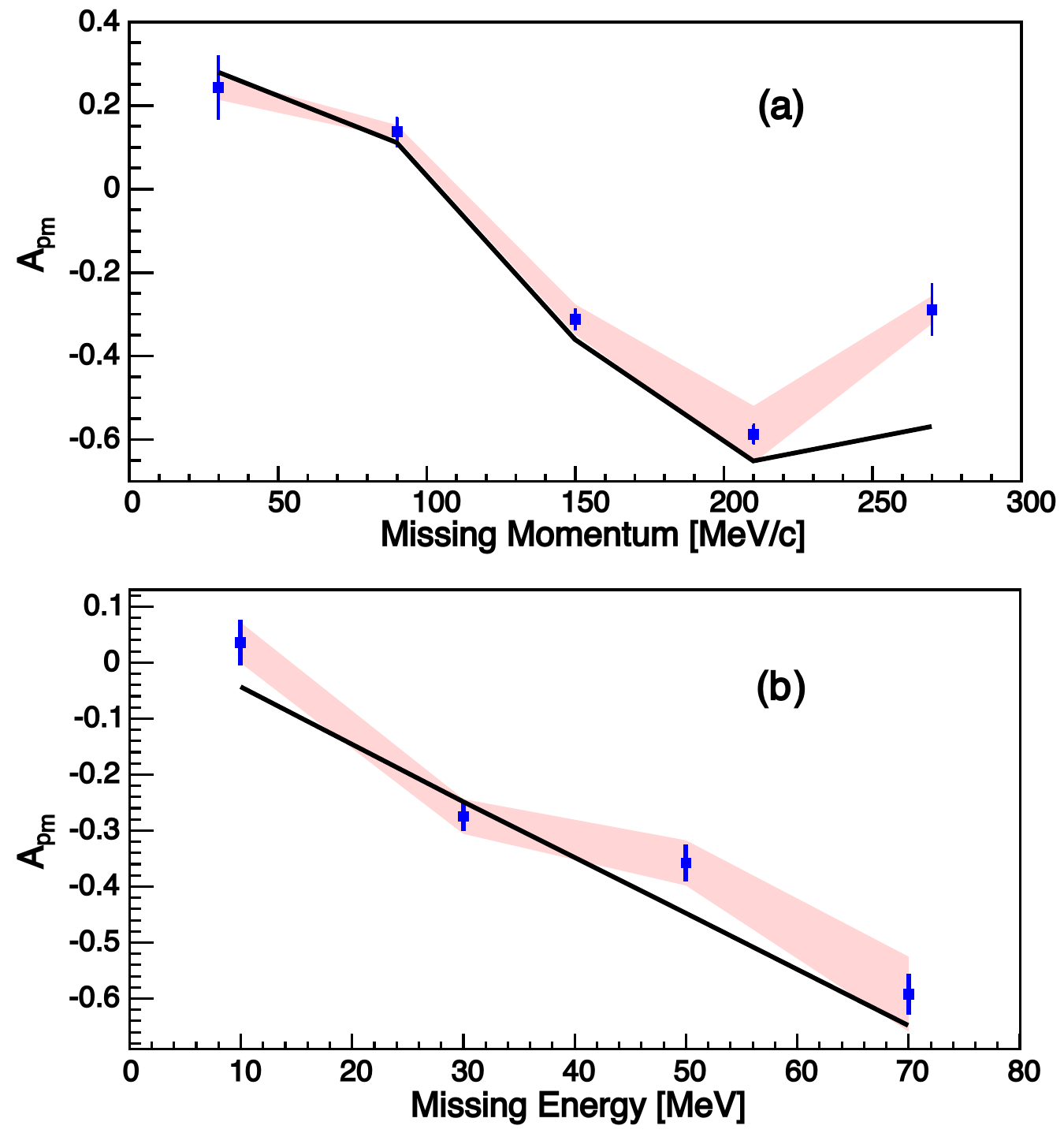}
 	     \caption{The measured missing momentum asymmetry as a function of the missing momentum (a) and missing energy (b) for $Q^2$=\,8.0\,(GeV/$c)^2$. The band shows the total systematic uncertainty which is the quadrature sum of the 11\% uncertainty introduced by the binning in $p_m$ and $E_m$ and the 4\% uncertainty from all the other sources listed in Table.~\ref{errors}. The black line indicates the simulated values for the corresponding points. }
 \label{fig: A_LT_ratio_missEP}
 \end{figure}

This is illustrated in Fig.~\ref{fig: A_LT_ratio_missEP} which shows the calculated $A_{p_m}$ as a function of the missing momentum and the missing energy for the $Q^2$ = 8.0~(GeV$/c)^2$ kinematic setting. The increase of $|A_{p_m}|$ with respect to $E_m$ and $|p_m|$ is as expected from the PWIA simulation (solid red lines). The small deviation at the highest missing momentum bin may be due to MEC that are not included in the simulation~\cite{PhysRevC.70.034606}. 

In the presence of additional FSI, such as when measuring in perpendicular kinematics, $|A_{p_m}|$ is known to decrease significantly relative to the PWIA expectation with increasing $E_m$ and $|p_m|$~\cite{PhysRevLett.83.5451}. Thus, measurements of $|A_{p_m}|$ in perpendicular kinematics could prove to be better probes of CT in future experiments. The signature of CT in such an experiment would be an increase in $|A_{p_m}|$ as a function of $Q^2$.\\
 \begin{figure}[h!]
\centering
\includegraphics[width=0.48\textwidth] 
{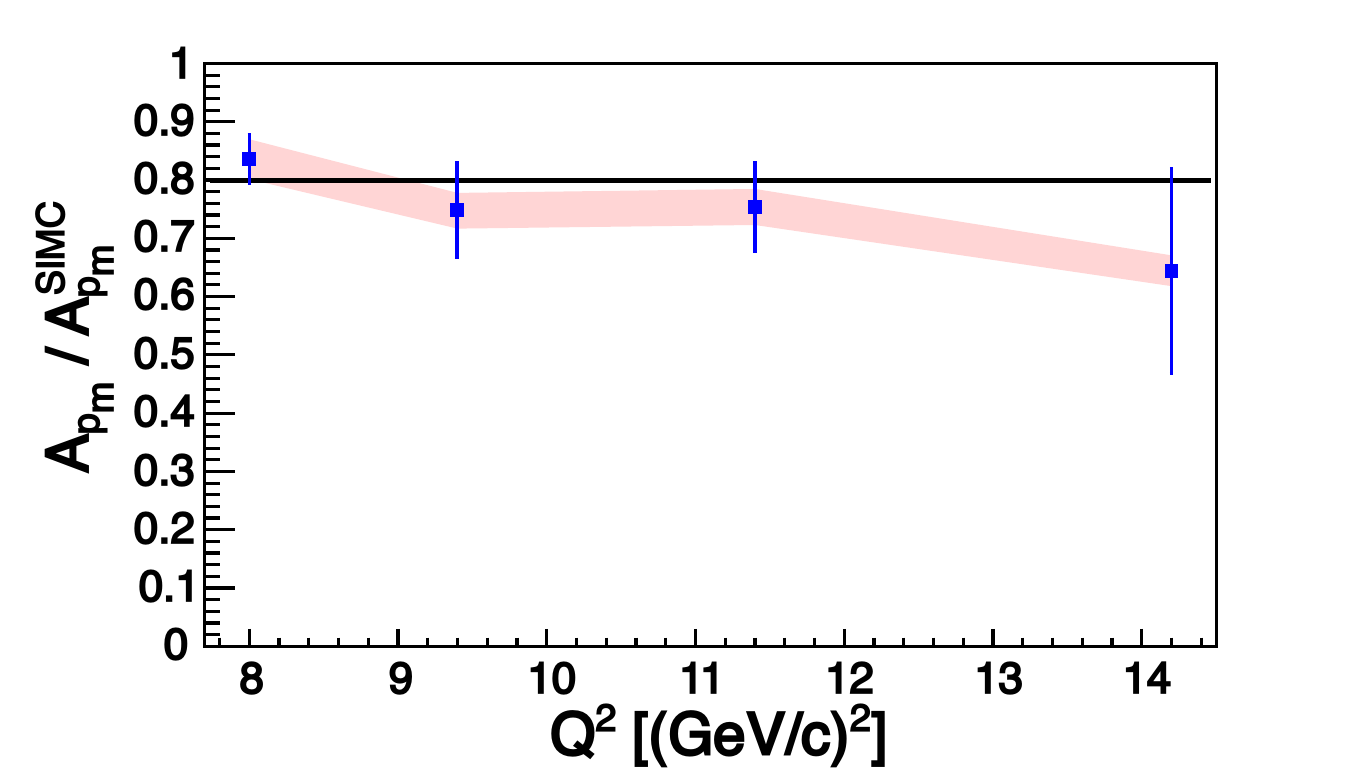}
 	     \caption{The ratio of the $A_{p_m}$ asymmetry in data to simulation as a function of $Q^2$. The band shows the total systematic uncertainty. The black line is the constant value fit to the data.}
 \label{fig: A_LT_ratio}
 \end{figure}
\indent Finally, Fig. \ref{fig: A_LT_ratio} shows the ratio of the measured $A_{p_m}$ asymmetry to the calculated asymmetry from the PWIA simulation as a function of $Q^2$. A range of $|p_m| \leq$~300 MeV/c and $E_m \leq$~80 MeV was used to extract the $A_{p_m}$ for all four $Q^2$ settings. The systematic uncertainty determined from varying the $E_m$ and $p_m$ range is $<$ 1\%, similar to what was observed for the transparency results. The $Q^2$ independence of the ratio indicates good agreement between the data and the PWIA simulation. The agreement between the measured and PWIA values of $A_{p_m}$ in parallel kinematics indicates the lack of CT-like effects or any additional FSI beyond the impulse approximation for the kinematics probed in this experiment.

\section{\label{sec:conclusions}CONCLUSIONS}
Using the upgraded 12\,GeV CEBAF beam at JLab, coincidence $(e,e'p)$ data were collected with $^{1}H$ and $^{12}C$ targets for $Q^2$ values between 8 and 14.2\,(GeV/$c)^2$.
The nuclear transparency was extracted at each of the four kinematic settings by integrating the charge-normalized yields and taking their ratio to the yields from a PWIA simulation of the experiment.
The transparency measured at the lowest kinematic point at $Q^2$=\,8\,(GeV/$c)^2$ agrees with prior measurements at JLab.
The $Q^2$ independence of the measured transparencies is consistent with traditional Glauber multiple scattering theory and does not show an onset of color transparency in $^{12}C(e,e'p)$ below $Q^2$=\,14.2\,(GeV/$c)^2$. We have also extracted the nuclear transparency of the $1s_{1/2}$ and $1p_{3/2}$ shell protons in $^{12}$C and their ratio. These observables show a $Q^2$ independence that rules out observation of the onset of CT for protons up to $Q^2$ of 14.2~(GeV$/c)^2$ in $^{12}C(e,e'p)$. We have also extracted the asymmetry of the  $^{12}C(e,e'p)$ events along and opposite to the momentum transfer $\vec{q}$ in parallel kinematics. The measured asymmetry is consistent with the expectations from a PWIA simulation of the experiment. These results rule out any additional reaction mechanisms such as CT for $^{12}C(e,e'p)$ in parallel kinematics.

\section{ACKNOWLEDGMENTS\\}

This work was funded in part by the U.S. Department of Energy, including contract 
AC05-06OR23177 under which Jefferson Science Associates, LLC operates Thomas Jefferson National Accelerator Facility, and by the U.S. National Science Foundation and the Natural Sciences and Engineering Research Council of Canada. We wish to thank the staff of Jefferson Lab for their vital support throughout the experiment. We are also grateful to all granting agencies providing funding support to authors throughout this project.

\bibliographystyle{apsrev4-1}
\bibliography{bibliography}

\end{document}